\documentclass[12pt]{article}

\usepackage{mathrsfs}
\usepackage[T1]{fontenc}
\usepackage{mathpazo}
\usepackage{setspace}
\usepackage{amsfonts}
\usepackage{amssymb}
\usepackage{amsmath}
\usepackage{esint}                   
\usepackage{epsfig}
\usepackage{latexsym}
\usepackage{color}
\usepackage{graphicx}
\usepackage[hidelinks]{hyperref}
\usepackage{nicefrac}
\usepackage[latin1]{inputenc}
\usepackage{pstricks}
\usepackage{slashed}
\usepackage{multirow}
\usepackage{soul}
\usepackage[nosort]{cite}

\def\hybrid{\topmargin -30pt    \oddsidemargin 0pt 
        \headheight 0pt \headsep 0pt
        \textwidth 6.25in       
        \textheight 9.5in       
        \marginparwidth .875in
        \parskip 5pt plus 1pt   \jot = 1.5ex}

\hybrid

\def\baselinestretch{1.2}

\catcode`\@=11

\def\marginnote#1{}
\newcount\hour
\newcount\minute
\newtoks\amorpm
\hour=\time\divide\hour by60
\minute=\time{\multiply\hour by60 \global\advance\minute by-\hour}
\edef\standardtime{{\ifnum\hour<12 \global\amorpm={am}%
        \else\global\amorpm={pm}\advance\hour by-12 \fi
        \ifnum\hour=0 \hour=12 \fi
        \number\hour:\ifnum\minute<10 0\fi\number\minute\the\amorpm}}
\edef\militarytime{\number\hour:\ifnum\minute<10 0\fi\number\minute}

\def\draftlabel#1{{\@bsphack\if@filesw {\let\thepage\relax
   \xdef\@gtempa{\write\@auxout{\string
      \newlabel{#1}{{\@currentlabel}{\thepage}}}}}\@gtempa
   \if@nobreak \ifvmode\nobreak\fi\fi\fi\@esphack}
        \gdef\@eqnlabel{#1}}
\def\@eqnlabel{}
\def\@vacuum{}
\def\draftmarginnote#1{\marginpar{\raggedright\scriptsize\tt#1}}

\def\draft{\oddsidemargin -.5truein
        \def\@oddfoot{\sl preliminary draft \hfil
        \rm\thepage\hfil\sl\today\quad\militarytime}
        \let\@evenfoot\@oddfoot \overfullrule 3pt
        \let\label=\draftlabel
        \let\marginnote=\draftmarginnote
   \def\@eqnnum{(\theequation)\rlap{\kern\marginparsep\tt\@eqnlabel}%
\global\let\@eqnlabel\@vacuum}  }

\def\draft2{
        \def\@oddfoot{\sl preliminary draft \hfil
        \rm\thepage\hfil\sl\today\quad\militarytime}
        \let\@evenfoot\@oddfoot \overfullrule 3pt
        \let\label=\draftlabel
        \let\marginnote=\draftmarginnote
   \def\@eqnnum{(\theequation)\rlap{\kern\marginparsep\tt\@eqnlabel}%
\global\let\@eqnlabel\@vacuum}  }

\def\preprint{\twocolumn\sloppy\flushbottom\parindent 2em
        \leftmargini 2em\leftmarginv .5em\leftmarginvi .5em
        \oddsidemargin -.5in    \evensidemargin -.5in
        \columnsep .4in \footheight 0pt
        \textwidth 10.in        \topmargin  -.4in
        \headheight 12pt \topskip .4in
        \textheight 6.9in \footskip 0pt
        \def\@oddhead{\thepage\hfil\addtocounter{page}{1}\thepage}
        \let\@evenhead\@oddhead \def\@oddfoot{} \def\@evenfoot{} }

\def\numberbysection{\@addtoreset{equation}{section}
        \def\theequation{\thesection.\arabic{equation}}}

\def\underline#1{\relax\ifmmode\@@underline#1\else
        $\@@underline{\hbox{#1}}$\relax\fi}

\def\titlepage{\@restonecolfalse\if@twocolumn\@restonecoltrue\onecolumn
     \else \newpage \fi \thispagestyle{empty}\c@page\z@
        \def\thefootnote{\fnsymbol{footnote}} }

\def\endtitlepage{\if@restonecol\twocolumn \else \newpage \fi
        \def\thefootnote{\arabic{footnote}}
        \setcounter{footnote}{0}}  

\catcode`@=12
\relax

\def\figcap{\section*{Figure Captions\markboth
        {FIGURECAPTIONS}{FIGURECAPTIONS}}\list
        {Figure \arabic{enumi}:\hfill}{\settowidth\labelwidth{Figure
999:}
        \leftmargin\labelwidth
        \advance\leftmargin\labelsep\usecounter{enumi}}}
 \relax
\def\tablecap{\section*{Table Captions\markboth
        {TABLECAPTIONS}{TABLECAPTIONS}}\list
        {Table \arabic{enumi}:\hfill}{\settowidth\labelwidth{Table
999:}
        \leftmargin\labelwidth
        \advance\leftmargin\labelsep\usecounter{enumi}}}
 \relax
\def\reflist{\section*{References\markboth
        {REFLIST}{REFLIST}}\list
        {[\arabic{enumi}]\hfill}{\settowidth\labelwidth{[999]}
        \leftmargin\labelwidth
        \advance\leftmargin\labelsep\usecounter{enumi}}}
 \relax
\makeatletter
\newcounter{pubctr}
\def\publist{\@ifnextchar[{\@publist}{\@@publist}}
\def\@publist[#1]{\list
        {[\arabic{pubctr}]\hfill}{\settowidth\labelwidth{[999]}
        \leftmargin\labelwidth
        \advance\leftmargin\labelsep
        \@nmbrlisttrue\def\@listctr{pubctr}
        \setcounter{pubctr}{#1}\addtocounter{pubctr}{-1}}}
\def\@@publist{\list
        {[\arabic{pubctr}]\hfill}{\settowidth\labelwidth{[999]}
        \leftmargin\labelwidth
        \advance\leftmargin\labelsep
        \@nmbrlisttrue\def\@listctr{pubctr}}}
 \relax
\makeatother

\def\be{\begin{equation}}
\def\ee{\end{equation}}
\def\ba{\begin{eqnarray}}
\def\ea{\end{eqnarray}}

\def\tr{\textrm{tr}}
\def\bg{\bar{g}}

\def\k{\kappa}
\def\r{\rho}
\def\a{\alpha}

\def\b{\beta}

\def\d{\delta}
\def\D{\Delta}

\def\th{\theta}

\def\m{\mu}
\def\n{\nu}

\def\l{\lambda}

\def\s{\sigma}

\def\cA{{\cal A}}
\def\cB{{\cal B}}

\def\cG{{\cal G}}
\def\cH{{\cal H}}
\def\cL{{\cal L}}

\def\cM{{\cal M}}

\def\cO{{\cal O}}

\def\no{\noindent}

\def\IR{\relax{\rm I\kern-.18em R}}

\def\inv{^{\raise.0ex\hbox{${\scriptscriptstyle -}$}\kern-.05em 1}}

\begin{document}


\renewcommand{\theequation}{\thesection.\arabic{equation}}
\csname @addtoreset\endcsname{equation}{section}

\begin{titlepage}
\begin{center}

\hfill HU-EP-25/24

\phantom{xxx}
\vskip 0.5in

{\large \bf Scale without conformal invariance\\ from integrable deformations of coset CFTs}

\vskip 0.5in

{\bf Georgios Itsios}${}^{1a}$\phantom{x}   and \phantom{x} {\bf Konstantinos Siampos}${}^{2b}$ \vskip 0.1in

${}^1$ Institut f\"{u}r Physik, Humboldt-Universit\"{a}t zu Berlin,\\
IRIS Geb\"{a}ude, Zum Gro{\ss}en Windkanal 2, 12489 Berlin, Germany\\

\vskip .2in

${}^2$  Laboratory of Theoretical Physics, School of Physics,\\
Aristotle University of Thessaloniki, 54124 Thessaloniki, Greece

\vskip .2in


\end{center}

\vskip .4in

\centerline{\bf Abstract}

\no 
We construct the integrable $\lambda$-model on $SU(3)_k/U(2)_k$ and we compute the one-loop $\beta$-function for the deformation parameter $\lambda$. Its non-compact version, corresponding to the coset $SU(2,1)_{-k}/U(2)_{-k}$ is also considered. The target space of the latter admits an asymptotic region that can be reached when one of the coordinates becomes large. The asymptotic model can be seen as an integrable deformation of the $SU(2)_k/U(1)$ WZW model together with a linear dilaton and a free boson, where integrability is inherited from the parental $\lambda$-model. By taking an asymptotic double-scaling limit of the $SU(2,1)_{-k}/U(2)_{-k}$ model, we obtain a two-dimensional field theory that is scale-invariant but not conformally invariant at one-loop order. Crucially, this deformation does not admit a Kerr-Schild form, unlike the cases studied in arXiv:2109.05040. However, in a suitable asymptotic regime, scale invariance is enhanced to full conformal invariance. Finally, we construct Type-II supergravity embeddings of the asymptotic model for specific values of the deformation parameter.

\vfill
\no
 {
$^a$georgios.itsios@physik.hu-berlin.de,  
$^b$ksiampos@auth.gr
}

\end{titlepage}
\vfill
\eject



\def\baselinestretch{1.2}
\baselineskip 20 pt

\newcommand{\eqn}[1]{(\ref{#1})}

\tableofcontents


\section{Introduction}

Relativistic field theories often exhibit additional symmetries, including scale invariance. A long-standing question in this context is whether the presence of scale symmetry necessarily implies conformal invariance. Typically, a field theory defined on flat spacetime and invariant under translations and rotations admits a conserved and symmetric energy-momentum tensor $T_{\mu\nu}$ (up to improvement terms). Scale invariance imposes further constraints, requiring the trace of $T_{\mu\nu}$ to take the so-called Virial form, i.e. $T^{\mu}{}_{\mu} = \partial_{\mu} V^{\mu}$, where $V^{\mu}$ is the Virial current. If the theory is also invariant under special conformal transformations, then $V^{\mu}$ must itself be a total divergence, i.e. $V^{\mu} = \partial_{\nu} Z^{\mu\nu}$, where $Z^{\mu\nu}$ is a symmetric rank-$2$ tensor \cite{Wess:1960,Gross:1970tb}. The condition for a conformally invariant theory is that $T_{\mu\nu}$ is traceless. However, scale invariance does not generically imply conformal invariance. The simplest known example where this happens is Maxwell's theory in $d$ spacetime dimensions \cite{El-Showk:2011xbs} (see also \cite{Gimenez-Grau:2023lpz,Papadopoulos:2024uvi,Chawla:2025fpn,Nakayama:2024jwq} for additional examples and \cite{Nakayama:2013is} for a review), where the trace of the energy-momentum tensor is
\begin{equation}
T^{\mu}{}_{\mu} =- \frac{d - 4}{2} \, \partial_{\mu} \big( A_{\nu} F^{\mu\nu} \big) \, .
\end{equation}
This expression indicates that Maxwell theory is scale invariant in arbitrary dimensions, but conformal invariance only emerges in $d = 4$, where the trace vanishes identically.

In two dimensions, Polchinski's theorem \cite{Polchinski.Scale}, based on an earlier work by Zamolodchikov \cite{Zamolodchikov:1986gt}, suggests that scale invariance is extended to conformal under the assumptions of (a) unitarity, (b) discrete spectrum of operator dimensions, and (c) the existence of energy-momentum tensor and its two-point functions. However, in the context of $\sigma$-models scale does not imply conformal invariance beyond tree-level if the diffeomorphism vector cannot be expressed as a derivative of a scalar (dilaton) field \cite{Hull:1985rc}. Examples that circumvent Polchinski's no-go theorem at one loop have been constructed in \cite{Itsios:2021eig}. These involve non-linear $\s$-models with non-compact target spaces, describing deformations of conformal field theories by parafermionic bilinears, and are based on orthogonal groups \cite{Sfetsos:2013wia}. Such models belong to a broader class of $\s$-models known as $\l$-deformed coset $\sigma$-models, which are integrable for a symmetric coset target space (both weakly and strongly) \cite{Hollowood:2014rla,Hollowood:2015dpa} and one-loop renormalisable \cite{Sfetsos:2014jfa,Appadu:2015nfa}. \footnote{The (integrable) $\lambda$-deformed (symmetric) coset $\sigma$-model interpolates between a gauged WZW and the non-Abelian T-dual of the (symmetric) coset Principal Chiral Model (PCM)~\cite{Sfetsos:2013wia}. For a symmetric target space, the corresponding PCM is classically integrable \cite{Eichenherr:1979ci}, as is its non-Abelian T-dual. The two are connected by a canonical transformation \cite{Curtright:1994be} and share the same equations of motion. In passing we note that there exist examples with a non-symmetric target space preserving integrability, such as the $U(3)/U(1)^3$ flat manifold \cite{Bykov:2014efa}.} The scale-invariant examples in \cite{Itsios:2021eig} arise through a limiting procedure involving the deformation parameter $\l$ and certain non-compact directions of the target space.

In the present work, we extend the approach of \cite{Itsios:2021eig} to $\l$-deformed coset CFTs based on unitary groups. Specifically, we investigate deformations of the gauged WZW model on $SU(3)_k/U(2)_k$ and its non-compact analogue on $SU(2,1)_{-k}/U(2)_{-k}$. By exploring various asymptotic limits of the non-compact model, we identify integrable deformations that are scale but non-conformal invariant. Notably, these deformations differ in nature from the cases examined in \cite{Itsios:2021eig}, as they cannot be expressed via a Kerr-Schild ansatz.

The paper is organised as follows: In Section~\ref{Sec:2}, we construct the $\l$-models on $SU(3)_k/U(2)_k$ and $SU(2,1)_{-k}/U(2)_{-k}$. To demonstrate their renormalisability at one loop, we compute the $\beta$-function for the deformation parameter $\l$. In Section~\ref{AsymptoticModel}, we analyse an asymptotic region of the $\l$-deformed $SU(2,1)_{-k}/U(2)_{-k}$ coset CFT, which can be interpreted as an integrable deformation of the $SU(2)_k/U(1)$ gauged WZW model together with a linear dilaton and a free boson. Section~\ref{ScaleInvariantModel} is devoted to the study of a double-scaling asymptotic limit of the $SU(2,1)_{-k}/U(2)_{-k}$ model, leading to a classically integrable two-dimensional field theory that is scale-invariant but not conformally invariant at one loop. We also identify a specific asymptotic limit in which conformal invariance is restored. Finally, in Section~\ref{Sec:SUGRA}, we construct Type-II supergravity embeddings of the asymptotic model for particular values of the deformation parameter $\l$.


\section{The $\l$-models on $SU(3)_k/U(2)_k$ and $SU(2,1)_{-k}/U(2)_{-k}$}
\label{Sec:2}

In this section we construct the $\l$-deformation \cite{Sfetsos:2013wia} of the $SU(3)_k/U(2)_k$ coset CFT. This is based on the WZW model that corresponds to the $SU(3)_k/U(2)_k$ coset. We also consider the deformation of the $SU(2,1)_{-k}/U(2)_{-k}$ coset CFT, whose corresponding WZW model was first constructed in \cite{Lugo:1994ra}. 


\subsection{Basics of $\l$-models}

Let us start by reviewing the $\l$-model for a semi-simple group $\cG$, developed in~\cite{Sfetsos:2013wia}. This is described by a two-dimensional action which looks like
\footnote{
When extracting background fields from the $\sigma$-model action, we assume that it takes the following form
\[
S = \frac{1}{2 \pi} \int d^2\xi \,  \big( G_{\m\n} + B_{\m\n} \big) \,\partial_+ x^\m\, \partial_- x^\n \,,
\]
with
\[
 d^2 \xi := d\tau \, d\sigma \, , \quad \sigma^\pm=\tau\pm\sigma\,,\quad
  \partial_\pm := \frac{1}{2} \big( \partial_\tau \pm \partial_\sigma \big) \, .
\]
\label{Normalization.action}
}

\begin{equation}
 \label{LambdaModelAction}
 S_{k,\l}[g] = S_{k, \text{WZW}}[g] - \frac{k}{\pi} \int d^2\xi \, R^A_{+} \, M^{-1}_{AB} \, L^B_{-} \, , \quad
 M := D^T - \l^{-1} \, .
\end{equation}
Here, $L^A_{\pm}$ and $R^A_{\pm}$ with $A = 1, \ldots , \text{dim}(\cG)$ stand for the pull-back of the left- and right-invariant one-forms on the world-sheet
\begin{equation}
 g^{-1} \, \partial_{\pm} g = i L^A_{\pm} \, t_A \, , \quad \partial_{\pm} g \, g^{-1} = i R^A_{\pm} \, t_A \, , \quad g \in \cG \, .
\end{equation}
Also, the matrix $D$ represents the adjoint action of the group
\begin{equation}
 D_{AB} = \tr \big( t_A g t_B g^{-1}  \big) \, , \quad g \in \cG
\end{equation}
and the Hermitian generators $t_A, \, (A = 1, \ldots , \textrm{dim}(\cG))$ are normalised such that $\tr\big(t_A t_B\big) = \d_{AB}$. As a result, the matrix $D$ is orthogonal. The deformation is given in terms of the $\textrm{dim}(\cG) \times \textrm{dim}(\cG)$ matrix $\l_{AB}$. Finally, $S_{k, \text{WZW}}[g]$ is the WZW action at level $k$, evaluated on an element $g \in \cG$
\begin{equation}
 \label{WZWAction}
 S_{k, \text{WZW}}[g] = \frac{k}{2 \pi} \int_{\partial\cM} d^2\xi \, L^A_{+} L^A_{-} + \frac{k}{24 \pi} \int_{\cM} d^3\xi \, f_{ABC} \, L^A \wedge L^B \wedge L^C \, ,
\end{equation}
with $f_{ABC}$ being the structure (real) constants of the group
\begin{equation}
 \big[ t_A , t_B \big] = i f_{AB}{}^C \, t_C \, , \qquad f_{ABC} := f_{AB}{}^E \, \d_{EC} \, .
\end{equation}
The $\s$-model with action \eqref{LambdaModelAction} is defined on a target space with metric
\begin{equation}
 \label{MetricLambda}
 ds^2 = k \, L^T \big( D - \l \big)^{-1} \big( \mathbb{1} - \l \l^T \big) \big( D - \l \big)^{- T} L \,,
\end{equation}
where $\mathbb{1}$ is the identity matrix of dimension $\textrm{dim}(\cG)$.
In general, one also obtains an antisymmetric field which is given in terms of the two-form below
\begin{equation}
 \label{BfieldLambda}
 B = B_\text{WZW} - \frac{k}{2} L^T \Big[ \big( \mathbb{1} - \l^{-1} D \big)^{-1} - \big( \mathbb{1} - \l^{-1} D \big)^{- T} \Big] \wedge L \, ,
\end{equation}
where $B_\text{WZW}$ corresponds to the antisymmetric field that arises from the WZ (second) term of \eqref{WZWAction}. On top of the above fields, one also finds a scalar $\Phi$ which reads
\begin{equation}
 \label{PhiLambda}
 e^{- 2 \Phi} = k^{\textrm{dim}(\cG)} \det M \, .
\end{equation}
This is obtained from the integration of non-dynamical gauge fields during the gauging procedure of the construction of the $\l$-model.

Notice that the action \eqref{LambdaModelAction} remains invariant under the map
\begin{equation}
 \label{LambdaModelSymmetry}
 \l \mapsto \l^{-1} \, , \quad k \mapsto - k \, , \quad g \mapsto g^{-1} \, .
\end{equation}
This symmetry must be reflected in the physical quantities derived from \eqref{LambdaModelAction}.

Another interesting feature of the $\l$-model is that it accommodates the non-Abelian T-dual (NATD) of a PCM. This becomes manifest in the limit where $\l \to \mathbb{1}$. In this case the action \eqref{LambdaModelAction} is singular and the limit makes sense if we adopt the following expansion for large values of $k$
\begin{equation}
\lambda=\mathbb{1}-\frac{E}{k}+{\cal O}(k^{-2})\,,\quad g=\mathbb{1}+\frac{v_At_A}{k}+{\cal O}(k^{-2})\,,\quad k\gg1 \, .
\end{equation}
Here $v_A$'s are background fields and $E$ is a generic $\text{dim}(\cG)$ square matrix. The above limit leads to
\begin{equation}
S_\text{NATD}=\frac1\pi\int d^2\xi\left( E + F \right)_{AB}^{-1}\partial_+v_A\partial_-v_B \, , \quad F_{AB} := f_{AB}{}^C \, v_C \, .
\end{equation}
This is recognized as the non-Abelian T-dual of the PCM
\begin{equation}
S_\text{PCM}=\frac1\pi\int d^2\xi\, E_{AB}R^A_+R^B_-\,,
\end{equation}
with respect to the right action of the group $\cG$.

Moving to deformations of cosets $\cG/\cH$, these are again described by \eqref{LambdaModelAction}. However, the $\l$ matrix now has a particular form, which follows from the split of the group $\cG$ generators into those of the $\cH$-subalgebra and its complement. More precisely, we take
\begin{equation}
 \label{generatorsSplit}
 t_A = (t_\a, \, t_a)
\end{equation}
where the lowercase Latin indices label the $\textrm{dim}(\cH)$ generators of the subgroup, and the lowercase Greek indices label the $\textrm{dim}(\cG/\cH)$ generators of the complement. If we now choose a deformation matrix of the form
\begin{equation}
 \label{LambdaCoset}
 \l_{AB} =
 \begin{pmatrix}
  \l \, \d_{\a\b} & \vline &  \mathbf{0}
  \\ \hline
  \mathbf{0} & \vline & \d_{ab}
 \end{pmatrix} \, ,
\end{equation}
the action \eqref{LambdaModelAction} develops a local $\cH$-invariance. This suggests that the $\l$-model now has $\textrm{dim}(\cH)$ redundant degrees of freedom, which can be fixed by an appropriate gauge choice of the group element $g$. The corresponding background fields are still given by \eqref{MetricLambda}, \eqref{BfieldLambda} and \eqref{PhiLambda}, where now the deformation matrix is \eqref{LambdaCoset}.

\subsubsection*{Integrability}

For a symmetric target space the corresponding $\l$-deformed model is classically integrable in the weak sense and its equations of motion can be expressed in terms of the Lax connection~\cite{Hollowood:2014rla}. The components of the latter are
\be
\label{Lax.Eq}
\cL_\pm=\cA_\pm+\frac{z^{\pm1}}{\sqrt{\l}}\cB_\pm\,,
\ee
where $z$ is the complex spectral parameter and
\be
\label{Lax.Eq.Quant}
\begin{split}
&\cA^a_+=i\,\left(\mathbb{1}-D\lambda^T\right)^{-1}_{a A}R^A_+\,,\quad
\cA^a_-=-i\left(\mathbb{1}-D^T\lambda\right)^{-1}_{a A}L^A_-\,,\\
&\cB^\alpha_+=i\lambda\left(\mathbb{1}-D\lambda^T\right)^{-1}_{\alpha A}R^A_+\,,\quad
\cB^\alpha_-=-i\lambda\left(\mathbb{1}-D^T\lambda\right)^{-1}_{\alpha A}L^A_- \, .
\end{split}
\ee
Here $\cA_\pm=\cA^a_\pm\,t_a$ belong in the algebra of the subgroup $\cH$, while $\cB_\pm=\cB^\a_\pm\,t_\a$ lie in its complement, in accordance with the decomposition \eqref{generatorsSplit}. The indices are raised and lowered using $\d_{AB}$, where $A=(\a,a)$. The explicit expressions of the gauge fields \( \cA_\pm \) and \( \cB_\pm \) are rather involved and will not be presented here. The Lax connection satisfies the flatness condition $\partial_+ \cL_- - \partial_- \cL_+ = [\cL_+, \cL_-]$, which ensures integrability in the weak sense. Moreover, the spatial component of the Lax connection assumes the Maillet form~\cite{Maillet2,Maillet}, which guarantees integrability in the strong sense~\cite{Hollowood:2015dpa}, where the corresponding conserved charges are in involution.

\subsubsection*{RG flow equations}

The $\l$-deformed coset model turns out to be one-loop renormalisable for symmetric target spaces \cite{Sfetsos:2014jfa,Appadu:2015nfa}. This can be inferred from the $\beta$-function (at one loop) of the metric \cite{Ecker:1971xko,Friedan:1980jm}
\begin{equation}
 \label{RGequation}
 \frac{d G_{\m\n}}{d\ln\mu^2} = R_{\m\n} + \nabla_\m \xi_\n + \nabla_\n \xi_\m \, ,
\end{equation}
where $R_{\m\n}$ is the Ricci tensor of the target space metric $G_{\m\n}$, $\xi=\xi^\mu\partial_\mu$ generates diffeomorphisms and $\mu$ is the 
energy scale. The corresponding RG flow equation for the deformation parameter is \cite{Sfetsos:2014jfa,Appadu:2015nfa} 
\be
\label{l.symmetric}
\frac{d\l}{d\ln\mu^2}=-\frac{c_\cG\l}{4k}\,,
\ee 
where $c_\cG$ is the quadratic Casimir of the group $\cG$ and $k$ is not running at one-loop order.


\subsection{Deforming the $SU(3)_k/U(2)_k$ coset CFT}
\label{TheLambdaModel}

We now continue with the deformation of the $SU(3)_k/U(2)_k$ symmetric coset CFT. As it was explained previously, we expect to have four redundant degrees of freedom, due to the local $U(2)$ symmetry.
\footnote{Before gauge fixing, the group element of $SU(3)$ is given by $H^\dagger g H$, where $H\in U(2)$ is given by~\cite{Lugo:1994ra} 
\begin{equation*}
H=h\oplus 1\,,\quad 
h=e^{\frac{i\delta}{2}(\mathbb{1}_2-\sigma_3)}\left( {\begin{array}{cc}
   n_1 & n_2 \\
   -n_2^* & n_1^* \\
  \end{array} } \right)\,,
\end{equation*}
with $|n_1|^2+|n_2|^2=1$.
}
This can be avoided by fixing the $SU(3)$ group element to be
\footnote{The inverse element $g^{-1}$ can be simply obtained by transposing $g$ in \eqref{GroupElement} and flipping the signs of $(\tau, \, \phi, \, x_1)$. That is
$
 g^{-1} = g^T\big|_{(\tau, \, \phi, \, x_1)\to(-\tau, \, -\phi, \,- x_1)} \, .
$
 Notice that the matrix parametrised by $x_0$ and $x_1$ can be seen also as a coset representative of $SU(2)/U(1)$. This becomes manifest by setting $x_0 = \cos\psi \cos\th$ and $x_1 = \sin\psi \cos\th$ and can also be written as
 \[
  g_{SU(2)/U(1)} = e^{\, i \frac{\psi}{2} \n_3} e^{\, i \th\, \n_2} e^{\, i \frac{\psi}{2} \n_3} \, .
 \]
}
\cite{Lugo:1994ra} 
\begin{equation}
 \label{GroupElement}
 g = e^{\, - i \tau \n_4} e^{\, i \frac{\sqrt{3}}{2} \phi \n_8}
  \begin{pmatrix}
   x_0 + i x_1                        & \sqrt{1 - x^2_0 - x^2_1}	&	0
   \\
   -\sqrt{1 - x^2_0 - x^2_1}   & x_0 - i x_1				&	0
   \\
   	0				 &		0				&	1
  \end{pmatrix} \, ,
\end{equation}
where the matrices $\n_i \, (i = 1, \ldots , 8)$ are the Gell-Mann matrices (see Appendix \ref{GellMannMatrices}) and 
on top of that, we take the $SU(3)$ generators to be
\begin{equation}
 t_A = \frac{1}{\sqrt{2}} \big( \n_4, \, \n_5, \, \n_6, \, \n_7, \, \n_1, \, \n_2, \, \n_3, \, \n_8 \big) \, .
\end{equation}
The generators $t_A$ above are normalised such that $\tr\big(t_A t_B\big) = \d_{AB}$ and the last four of them consist a $U(2)$ subalgebra. In this normalisation the quadratic Casimir of the $SU(3)$ equals to $c_\cG=6$.

The $\l$-deformation on $SU(3)_k/U(2)_k$ is derived by inserting the gauge fixed group element \eqref{GroupElement} inside \eqref{LambdaModelAction} and choosing the $\l$-matrix to be
\begin{equation}
 \l_{AB} =
 \begin{pmatrix}
  \l \, \mathbb{1}_4 & \vline &  \mathbf{0}
  \\ \hline
  \mathbf{0} & \vline & \mathbb{1}_4
 \end{pmatrix} \, .
\end{equation}
The corresponding $\s$-model is classically integrable having a four-dimensional symmetric target space, \footnote{The corresponding Lax pair can be found from \eqref{Lax.Eq} and \eqref{Lax.Eq.Quant} using the parameterization of \eqref{GroupElement}.} whose metric takes the following form
\begin{equation}
 \label{LambdaDeformedMetric}
 G_{\m\n} = 2 \, k \left( \frac{1 + \l^2}{1 - \l^2} \, \bg_{\m\n} + \frac{\l}{1 - \l^2} \, g_{\m\n}\right) \, .
\end{equation}
Here $G_{\m\n} \big|_{\l = 0} = 2 \, k \, \bg_{\m\n}$ stands for the target space metric of the WZW model on the coset $SU(3)_k/U(2)_k$ and $g_{\m\n}$ are the components generated by the deformation. Both $\bg_{\m\n}$ and $g_{\m\n}$ are explicitly given below. In particular, the non-vanishing entries of $\bg_{\m\n}$ are
\begin{equation}
 \label{LambdaZeroMetric}
 \begin{aligned}
  & \bg_{\tau\tau} = 1 \, , \qquad\qquad
     \bg_{x_0 x_0} =\frac{1}{\rho^2} \cot^2\frac{\tau}{2} \, , \qquad\qquad
     \bg_{x_1 x_1} = \frac{1}{\rho^2} \tan^2\frac{\tau}{2} \, ,
     \\[5pt]
  & \bg_{\phi\phi} = \cot^2\tau + \frac{x^2_0}{4\r^2} \tan^2\frac{\tau}{2} + \frac{x^2_1}{4\r^2} \cot^2\frac{\tau}{2} \, ,
     \\[5pt]
  & \bg_{\phi x_0} = \frac{x_1}{2  \r^2} \, \cot^2 \frac{\tau}{2} \, , \qquad\qquad
     \bg_{\phi x_1} = - \frac{x_0}{2  \r^2} \, \tan^2 \frac{\tau}{2} \, ,
 \end{aligned}
\end{equation}
where we defined the function $\r := \sqrt{1 - x^2_0 - x^2_1}$.
The coordinates $x_0$ and $x_1$ must take values such that the combination in the square root of the previous expression is non-negative. Moving to the components of $g_{\m\n}$ we find
\begin{equation}
 \label{MetricDef}
 \begin{aligned}
  & g_{\tau\tau} = 2 W_{-} \, , \quad\quad
     g_{\tau\phi} = x_0 \cot\frac{\tau}{2} \sin\frac{3 \phi}{2} - x_1 \tan\frac{\tau}{2} \cos\frac{3 \phi}{2} \, ,
  \\[5pt]
  & g_{\tau x_0} = - 2 \cot\frac{\tau}{2} \cos\frac{3 \phi}{2} \, , \quad\quad
     g_{\tau x_1} = - 2 \tan\frac{\tau}{2} \sin\frac{3 \phi}{2} \, ,
  \\[5pt]
  & g_{\phi\phi} = \frac{(1 - \r^2) W_{-}}{4 \r^2} \frac{\cos2 \tau}{\sin^2\tau} - \frac{(1 + \r^2) W_{+}}{\r^2} \frac{\cot\tau}{\sin\tau}
  \\[5pt]
  & \qquad + \frac{x_0 \big( 3 x^2_0 - 5 x^2_1 \big) \cos\frac{3 \phi}{2} - x_1 \big( 3 x^2_1 - 5 x^2_0 \big) \sin\frac{3 \phi}{2}}{4 \r^2 \sin^2\tau} \, ,
  \\[5pt]
  & g_{\phi x_0} = - \frac{x_0 x_1 \cos\frac{3 \phi}{2}}{\r^2 \sin^2 \frac{\tau}{2}} + \frac{\D \sin\frac{3 \phi}{2}}{2 \r^2 \sin^2\frac{\tau}{2}} \, , \quad
     g_{\phi x_1} = - \frac{x_0 x_1 \sin\frac{3 \phi}{2}}{\r^2 \cos^2 \frac{\tau}{2}} - \frac{\D \cos\frac{3 \phi}{2}}{2 \r^2 \cos^2\frac{\tau}{2}} \, ,
  \\[5pt]
  & g_{x_0 x_0} = -  \frac{2W_{+}}{\r^2} \cot^2\frac{\tau}{2} \, , \quad
     g_{x_0 x_1} =  \frac{2Q_{-}}{\r^2} \, , \quad
     g_{x_1 x_1} =  \frac{2W_{+}}{\r^2} \tan^2\frac{\tau}{2} \, ,
 \end{aligned}
\end{equation}
where for the sake of a better presentation we define
\begin{equation}
 \begin{aligned}
  & W_\pm := x_0 \cos\frac{3 \phi}{2} \pm x_1 \sin\frac{3 \phi}{2} \, , \quad 
      Q_\pm := x_1 \cos\frac{3 \phi}{2} \pm x_0 \sin\frac{3 \phi}{2} \, ,
     \\[5pt]
   & \D := x^2_0 - x^2_1 - (1 + \r^2) \cos\tau \, .
 \end{aligned}
\end{equation}
Clearly, the metric $G_{\m\n}$ given in \eqref{LambdaDeformedMetric}, \eqref{LambdaZeroMetric} and \eqref{MetricDef} respects the symmetry \eqref{LambdaModelSymmetry}, without altering the coordinates. In addition, the metric is invariant under the $\mathbb{Z}_2$ grading $\lambda\to-\lambda$ provided that the coordinates map as $(x_0,x_1)\mapsto(-x_0,-x_1)$ or $(x_0,\phi)\mapsto(-x_0,\pi+\phi)$. Regarding the rest of the background fields, it turns out that this $\s$-model has vanishing antisymmetric field but it is supported by a non-trivial scalar field $\Phi$ given below
\begin{equation}
 \label{LambdaScalar}
 e^{- 2 \Phi} = 3 k^8 \frac{(1 - \l^2)^2}{\l^4} \r^2 \sin^4\tau \, .
\end{equation}
As indicated from the dilaton \eqref{LambdaScalar} the target space metric $G_{\mu\nu}$ is singular (Ricci scalar blows up) at $\rho=0$, circle of radius one in the $(x_0,x_1)$-plane, and at $\tau=0$.
One can show that the above metric $G_{\m\n}$ and scalar field $\Phi$ satisfy the following equations
\begin{equation}
\begin{split}
\label{betaPhibetaG}
& \b_{\Phi} := R + 4 \, \nabla^2 \Phi - 4 \, \big( \partial \Phi \big)^2 = \frac{6}{k} \frac{1 + \l^2}{1 - \l^2} \,, \\[5pt]
& \b_{G_{\m\n}} := R_{\m\n} + 2 \nabla_\m \partial_\n \Phi =
-\frac{12\l^2}{(1-\l^2)^2}\bg_{\m\n}- \frac{3\l(1+\l^2)}{(1 - \l^2)^2} \, g_{\m\n} \, . 
\end{split}
\end{equation}
At the conformal point $\l = 0$, only $\b_\Phi$ is non-trivial taking the value $\nicefrac{6}{k}$.

\subsubsection*{RG flow equations}

The RG flow equations \eqref{RGequation} for the background \eqref{LambdaDeformedMetric} can be solved by setting
\begin{equation}
 \xi_{\m} = \partial_{\m} \Phi \, .
\end{equation}
This translates into a flow for the deformation parameter, which in this case reads 
\begin{equation}
\label{RG.lambda.su3.su2}
 \frac{d \l}{d \ln \m^2} = - \frac{3\l}{2k} \, ,
\end{equation}
and is in agreement with \eqref{l.symmetric} for $c_\cG=6$. A comment is in order related to the latter expression. Consider a conformal field theory perturbed by operators of the form $\lambda_i \mathcal{O}_i$, where $\mathcal{O}_i$ are operators with scaling dimensions $\Delta_i$. The RG flow equations to two-loop order in perturbation theory are given by (see for example \cite{Brezin:1990sk})
\be
\frac{d\lambda_i}{d\ln\mu^2}=\frac12(\Delta_i-2)\lambda_i+\frac12C_i{}^{jk}\lambda_j\lambda_k+{\cal O}(\lambda^3) \, ,
\ee
where $C_i{}^{jk}$ are the fusion coefficients of the operators $\cO_i$. In the present case, the beta function is linear, indicating that the perturbing operator has scaling dimension $2 - \nicefrac{3}{k}$, corresponding to a deformation of the $SU(3)_k / U(2)_k$ coset CFT driven by parafermion bilinears.

Using \eqref{betaPhibetaG}, we may read the $C$-function of the model following~\cite{Tseytlin:1987bz,Tseytlin:2006ak,Sagkrioti:2018abh}. This is given by
\begin{equation}
\label{c-function.example}
C(\lambda,k)=4-3\beta_\Phi=4-\frac{18}{k}\frac{1+\lambda^2}{1-\lambda^2}+{\cal O}\left(\frac{1}{k^2}\right)\,,
\end{equation}
and is invariant under the symmetry~\eqref{LambdaModelSymmetry} as well as the $\mathbb{Z}_2$-grading $\lambda\to-\lambda$. For $\lambda=0$, we recover the central charge of the underlying UV CFT on $SU(3)_k/U(2)_k$,
\footnote{In our conventions, the central charge of a group $\cG$ at level $k$ is given by 
$
 c=\frac{2 \, k \, \text{dim}(\cG)}{2k+c_\cG}\,.
$}
\begin{equation}
\label{c.CFT}
c=\frac{8 k}{k+3}-\frac{3 k}{k+2}-1=4-\frac{18}{k}+{\cal O}\left(\frac{1}{k^2}\right)\,.
\end{equation}
Furthermore, the C-function \eqref{c-function.example} satisfies Zamolodchikov's c-theorem \cite{Zamolodchikov:1986gt} (see also \cite{Sagkrioti:2018abh} for this class of models)
\begin{equation}
 \frac{d C}{d \ln \m^2} = \frac{108 \lambda ^2}{k^2 \left(1-\lambda ^2\right)^2}\geqslant0\,.
\end{equation}


\subsubsection*{The NATD limit}

As it has already been observed below eq. \eqref{LambdaDeformedMetric}, when $\l = 0$ one recovers the WZW model on $SU(3)_k/U(2)_k$. This is one point of the interpolation, while the other is obtained when $\l$ approaches one \cite{Sfetsos:2013wia}. However, one cannot strictly set $\l = 1$ since the metric \eqref{LambdaDeformedMetric} diverges. Nevertheless, one can circumvent this via a correlated limit between the parameters $\l$ and $k$, which also involves a zoom-in of the group element $g$. In particular, as it was explained in \cite{Sfetsos:2013wia}, this can be achieved when $\l$ approaches one as $k$ becomes large, i.e.
\begin{equation}
 \label{expansionLambda}
\l = 1 - \frac{\kappa^2}{k} + \cO\left(\frac{\k^4}{k^2}\right) \, , \quad k \gg \k^2 \,,
\end{equation}
and at the same time $g$ approaches the identity element.

In our case, we will expand the $SU(3)$ group element \eqref{GroupElement} according to the following fashion
\begin{equation}
 g = \mathbb{1} + \frac{i \k^2}{k} \big( v_1 t_1 + v_6 t_6 + v_7 t_7 + v_8 t_8 \big) + \cO\left(\frac{\k^4}{k^2}\right) \, .
\end{equation}
In other words, this corresponds to expanding the group parameters (or $\s$-model coordinates) as
\begin{equation}
 \label{expansionGroupParameters}
 \begin{aligned}
  & \tau = - \frac{\kappa^2}{\sqrt{2}\, k} v_1 + \cO\left(\frac{\k^4}{k^2}\right) \, , \qquad\qquad\,\,\,\,
 \phi = \sqrt{\frac{2}{3}} \frac{\kappa^2}{k} v_8 +\cO\left(\frac{\k^4}{k^2}\right) \, ,
 \\[5pt]
 & x_0 = 1 - \kappa^4\frac{v^2_6 + v^2_7}{4 k^2} + \cO\left(\frac{\k^6}{k^3}\right) \, , \qquad
 x_1 = \frac{\kappa^2}{\sqrt{2} k} v_7 +\cO\left(\frac{\k^4}{k^2}\right)\, .
 \end{aligned}
\end{equation}

Inserting the expansions \eqref{expansionLambda} and \eqref{expansionGroupParameters} into the metric of the deformed $\s$-model \eqref{LambdaDeformedMetric}, \eqref{LambdaZeroMetric}, \eqref{MetricDef} and sending $k \to \infty$, one derives the metric for the NATD
\footnote{
 The NATD is applied on $SU(3)$.
}
of $\mathbb{CP}^2 \simeq SU(3)/U(2)$ \cite{Lozano:2011kb}. This can be conveniently expressed in terms of the vierbein below (as $ds^2 = \k^2 \d_{ij} \, e^i e^j$)
 \begin{align}
  & e^1 =  \frac{2}{\sqrt{3}} \frac{dv_8}{v_1} \, ,
  \nonumber\\[5pt]
  & e^2 =  \frac{\sqrt{2}}{v_1} \big( v_1 dv_1 + v_6 dv_6 + v_7 dv_7 + v_8 dv_8 \big) \, ,
  \\[5pt]
  & e^3 =  \frac{2}{v_1} dv_6 + \frac{2 v_7}{v_1 v_6} \Big( dv_7 - \frac{dv_8}{\sqrt{3}} \Big) \, ,
  \nonumber\\[5pt]
  & e^4 =  \sqrt{2} \big(v_7 - \sqrt{3} v_8 \big) \left( \frac{v_6}{v_1} dv_6 - \frac{v_7}{v_1 v_6} \Big( dv_7 - \frac{dv_8}{\sqrt{3}} \Big) \right)
  -  \sqrt{\frac{2}{3}} \frac{v_6}{v_1} dv_8 - \frac{v_1}{\sqrt{2} v_6} \Big( dv_7 - \frac{dv_8}{\sqrt{3}} \Big) \, .
  \nonumber
 \end{align}
Moreover, in this limit, the scalar $\Phi$ from \eqref{LambdaScalar} becomes
\begin{equation}
e^{- 2 \Phi} = \frac{3\kappa^8}{2} v^4_1 v^2_6 \, .
\end{equation}

The RG flow of the parameter $\kappa^2$ can be determined using \eqref{RGequation}
\begin{equation}
\frac{d\kappa^2}{dt} = \frac{3}{2}\,, \qquad 
\xi = G^{\mu\nu} \partial_\nu \Phi \, \partial_\mu
-\frac{3}{2\kappa^2} \left( v_1 \partial_{v_1} + v_6 \partial_{v_6} + v_7 \partial_{v_7} + v_8 \partial_{v_8} \right) \,,
\end{equation}
which is consistent with \eqref{RG.lambda.su3.su2} in the limit \eqref{expansionLambda}.

Notice that the NATD of $\mathbb{CP}^2$ that we found here, by taking the limit $\l \to 1$ in the $\l$-model of the $SU(3)_k/U(2)_k$ coset CFT, does not match the one found in \cite{Lozano:2011kb}. The reason is that the two cases correspond to different gauge choices of the Lagrange multipliers $v_i \, (i = 1, \ldots , 8)$. In particular, the NATD background found here corresponds to the gauge fixing given by setting $v_2 = v_3 = v_4 = v_5 = 0$, while the one of \cite{Lozano:2011kb} to the gauge choice $v_2 = v_4 = v_6 = v_7 = 0$.


\subsection{The $\l$-model on $SU(2,1)_{-k}/U(2)_{-k}$}
\label{TheNCLambdaModel}

We now continue with the integrable deformation of the $SU(2,1)_{-k}/U(2)_{-k}$ coset CFT. This can be easily obtained from the integrable $\l$-model on $SU(3)_k/U(2)_k$ by applying the analytic continuation $\tau \to i \tau$ together with $k \to - k$
\footnote{
 The last ensures that the metric has the correct signature.
}.
Such a transformation does not affect the integrability of the corresponding $\s$-model. Moreover, the target space is now non-compact, as $\tau$ is valued in $\mathbb{R}$. Its metric again takes the form \eqref{LambdaDeformedMetric}, where now
\begin{equation}
 \label{LambdaZeroMetricNC}
 \begin{aligned}
  & \bg_{\tau\tau} = 1 \, , \quad\quad
     \bg_{x_0 x_0} = \frac{\coth^2\frac{\tau}{2}}{\r^2} \, , \quad\quad
     \bg_{x_1 x_1} = \frac{\tanh^2\frac{\tau}{2}}{\r^2} \, ,
     \\[5pt]
  & \bg_{\phi\phi} = \frac{3 + \r^2 - 4 (x^2_0 - x^2_1) \cosh \tau + (1 + 3 \r^2) \cosh(2 \tau)}{8 \, \r^2 \sinh^2 \tau} \, ,
     \\[5pt]
  & \bg_{\phi x_0} = \frac{x_1}{2 \, \r^2} \, \coth^2 \frac{\tau}{2} \, , \quad\quad
     \bg_{\phi x_1} = - \frac{x_0}{2 \, \r^2} \, \tanh^2 \frac{\tau}{2} \, ,
 \end{aligned}
\end{equation}
and
\begin{equation}
 \label{MetricDefNC}
 \begin{aligned}
  & g_{\tau\tau} = 2 W_{-} \, , \quad\quad
     g_{\tau\phi} = - x_0 \coth\frac{\tau}{2} \sin\frac{3 \phi}{2} - x_1 \tanh\frac{\tau}{2} \cos\frac{3 \phi}{2} \, ,
  \\[5pt]
  & g_{\tau x_0} = 2 \coth\frac{\tau}{2} \cos\frac{3 \phi}{2} \, , \quad\quad
     g_{\tau x_1} = - 2 \tanh\frac{\tau}{2} \sin\frac{3 \phi}{2} \, ,
  \\[5pt]
  & g_{\phi\phi} = \frac{(1 - \r^2) W_{-}}{4 \r^2} \frac{\cosh(2 \tau)}{\sinh^2\tau} - \frac{(1 + \r^2) W_{+}}{\r^2} \frac{\coth\tau}{\sinh\tau}
  \\[5pt]
  & \quad + \frac{x_0 \big( 3 x^2_0 - 5 x^2_1 \big) \cos\frac{3 \phi}{2} - x_1 \big( 3 x^2_1 - 5 x^2_0 \big) \sin\frac{3 \phi}{2}}{4 \r^2 \sinh^2\tau} \, ,
  \\[5pt]
  & g_{\phi x_0} = - \frac{x_0 x_1 \cos\frac{3 \phi}{2}}{\r^2 \sinh^2 \frac{\tau}{2}} + \frac{\D \sin\frac{3 \phi}{2}}{2 \r^2 \sinh^2\frac{\tau}{2}} \, , \quad
     g_{\phi x_1} = \frac{x_0 x_1 \sin\frac{3 \phi}{2}}{\r^2 \cosh^2 \frac{\tau}{2}} + \frac{\D \cos\frac{3 \phi}{2}}{2 \r^2 \cosh^2\frac{\tau}{2}} \, ,
  \\[5pt]
  & g_{x_0 x_0} = - \frac{2W_{+}}{\r^2} \coth^2\frac{\tau}{2} \, , \quad
     g_{x_0 x_1} = -  \frac{2Q_{-}}{\r^2} \, , \quad
     g_{x_1 x_1} =  \frac{2W_{+}}{\r^2} \tanh^2\frac{\tau}{2} \, .
 \end{aligned}
\end{equation}
Notice that when $\l = 0$ the metric components define the target space metric of the WZW model on $SU(2,1)_{-k}/U(2)_{-k}$ constructed in \cite{Lugo:1994ra}. As it is expected, the metric $G_{\m\n}$ given in \eqref{LambdaDeformedMetric}, \eqref{LambdaZeroMetricNC} and \eqref{MetricDefNC} respects the symmetry \eqref{LambdaModelSymmetry}. 

In analogy to the compact case, this $\s$-model has vanishing antisymmetric field, but the scalar $\Phi$ is now given by
\begin{equation}
 \label{LambdaScalarNC}
 e^{- 2 \Phi} = 3 k^8 \frac{(1 - \l^2)^2}{\l^4} \r^2 \sinh^4\tau \, .
\end{equation}
Now the $\beta$-functions $\b_{\Phi}$ and $\b_{G_{\m\n}}$ defined in \eqref{betaPhibetaG} have the opposite sign, i.e.
\begin{eqnarray}
 \label{betaPhibetaGNC}
  && \b_{\Phi} = R + 4 \, \nabla^2 \Phi - 4 \, \big( \partial \Phi \big)^2 = - \frac{6}{k} \frac{1 + \l^2}{1 - \l^2} \, ,
  \\[5pt]
  && \b_{G_{\m\n}} = R_{\m\n} + 2 \nabla_\m \nabla_\n \Phi = \frac{3}{2 \, k} \, \frac{1 + \l^2}{1 - \l^2} G_{\m\n} - 3 \, \bg_{\m\n}=
\frac{12\l^2}{(1-\l^2)^2}\bg_{\m\n}+ \frac{3\l(1+\l^2)}{(1 - \l^2)^2} \, g_{\m\n} \, .\nonumber
\end{eqnarray}
This is expected since the derivation of the $\l$-model on $SU(2,1)_{-k}/U(2)_{-k}$ from that on $SU(3)_k/U(2)_k$ involves flipping the sign of $k$. For the same reason, the RG flow at one-loop for the deformation parameter now is
\begin{equation}
\label{RGs.oneloop.su12ovu2}
 \frac{d \l}{d \ln \m^2} = \frac{3 \, \l}{2 \, k} \, .
\end{equation}

Similarly to the $SU(3)_k/U(2)_k$ case, the $c$-function is given by \eqref{c-function.example} with the replacement $k \to -k$
\begin{equation}
\label{Cfunction.noncompact}
C(\lambda,k) = 4 + \frac{18}{k} \frac{1 + \lambda^2}{1 - \lambda^2} +{\cal O}\left(\frac{1}{k^2}\right)\,,
\end{equation}
which coincides at $\lambda = 0$ with the central charge of the underlying CFT on $SU(2,1)_{-k}/U(2)_{-k}$ (see \eqref{c.CFT}). It also obeys Zamolodchikov's $c$-theorem~\cite{Zamolodchikov:1986gt}.


\section{The asymptotic model}
\label{AsymptoticModel}

The $\l$-model on $SU(2,1)_{-k}/U(2)_{-k}$ described in Section \ref{TheNCLambdaModel} admits a non-compact target space and an asymptotic region that can be reached for large values of the coordinate $\tau$
\footnote{
Coset CFTs or deformations of them with the same feature have been studied in \cite{Lugo:1994ra,Petropoulos:2006py,Itsios:2021wso}
}. 
Indeed, when $\tau \gg 1$ the components \eqref{LambdaZeroMetricNC} become
\begin{equation}
 \label{LambdaZeroAsymMetric}
 \begin{aligned}
  & \bg_{\tau\tau} = 1 \, , \qquad\qquad
     \bg_{x_0 x_0} =  \bg_{x_1 x_1} = \frac{1}{\r^2} \, , \qquad\qquad
     \bg_{\phi\phi} = \frac{1 + 3 \r^2}{4 \, \r^2} \, ,
     \\[5pt]
  & \bg_{\phi x_0} = \frac{x_1}{2 \, \r^2} \, , \qquad\qquad
     \bg_{\phi x_1} = - \frac{x_0}{2 \, \r^2} \, .
 \end{aligned}
\end{equation}
Similarly, for the non-zero components \eqref{MetricDefNC} we find
\begin{equation}
 \begin{aligned}
  \label{AsymMetricDef}
  & g_{\tau\tau} = 2 W_- \, , \qquad
     g_{\tau\phi} = - Q_+ \, , \qquad
     g_{\tau x_0} = 2 \cos\frac{3 \phi}{2} \, , \qquad
     g_{\tau x_1} = - 2 \sin\frac{3 \phi}{2} \, ,
  \\[5pt]
   & g_{\phi\phi} = \frac{1 - \r^2}{2 \r^2} W_- \, , \qquad
  g_{\phi x_0} = - \frac{1 + \r^2}{\r^2} \sin\frac{3 \phi}{2} \, , \qquad
  g_{\phi x_1} =  - \frac{1 + \r^2}{\r^2} \cos\frac{3 \phi}{2}  \, ,
  \\[5pt]
  & g_{x_0 x_0} = - g_{x_1 x_1} = - \frac{2W_+}{\r^2}  \, , \qquad
     g_{x_0 x_1} = - \frac{2Q_-}{\r^2} \, .
 \end{aligned}
\end{equation}
The full metric is again given by the formula \eqref{LambdaDeformedMetric}, where now $\bg_{\m\n}$ and $g_{\m\n}$ take the above values. On the other hand, the leading contribution to the scalar \eqref{LambdaScalarNC} is
\begin{equation}
 \label{AsymScalar}
 e^{- 2 \Phi} = \frac{3}{16} k^8 \frac{(1 - \l^2)^2}{\l^4} e^{4 \tau} \r^2 \, .
\end{equation}
As it can be seen, the scalar $\Phi$ is obviously linear in $\tau$.

Like in the original $\l$-model one can show that the asymptotic target space metric together with with the scalar \eqref{AsymScalar} also satisfy the equations \eqref{betaPhibetaGNC}. The RG flow for the parameter $\l$ at one loop is still given by \eqref{RGs.oneloop.su12ovu2}. Moreover, one can express the equations of motion of the asymptotic $\s$-model in terms of a Lax connection demonstrating its integrability. This property has been inherited from the parental $\l$-model on $SU(2,1)_{-k}/U(2)_{-k}$.


\subsection{The $\l \to 0$ and $\l \to 1$ limits}

Let us now study the behaviour of the asymptotic model when $\l$ approaches the values of zero and one.

\noindent \textbf{The $\l \to 0$ limit:} Clearly, when $\l = 0$ the components of \eqref{AsymMetricDef} do not contribute to the target space metric $G_{\m\n}$. The line element in this case reads
\begin{equation}
 ds^2 = 2k \Bigg( d\tau^2 + \frac{dx^2_0 + dx^2_1}{\r^2} + \frac{x_1}{\r^2} \, dx_0 \, d\phi - \frac{x_0}{\r^2} \, dx_1 \, d\phi + \frac{1 + 3 \r^2}{4 \r^2} d\phi^2 \Bigg) \, .
\end{equation}
A more convenient parametrisation can be found by rotating $x_0$ and $x_1$ in the following fashion
\begin{equation}
 x_0 = \cos\th \, \cos \Big( \psi + \frac{\phi}{2} \Big) \, , \qquad
 x_1 = \cos\th \, \sin \Big( \psi + \frac{\phi}{2} \Big) \, .
\end{equation}
The line element now reads
\begin{equation}
 \label{Lambda0AsymptGeom}
 ds^2 = 2 k \big( d\th^2 + \cot^2\th \, d\psi^2+ d\tau^2+d\phi^2 \big)
\end{equation}
and as anticipated, the geometry is singular when $\th = 0$ where $R = - \frac{2}{k \, \sin^2\th}$. 
For the scalar of \eqref{AsymScalar} we find
\begin{equation}
 \label{Lambda0AsymptScalar}
 e^{- 2 \Phi} = \frac{3}{16} k^8 \, e^{4 \tau} \sin^2\th \,,
\end{equation}
which also blows up when $\th = 0$.

This field content corresponds to a CFT described by the gauged WZW model on $SU(2)_k/U(1)$, accompanied by a linear dilaton parametrized by $\tau$, and a free boson $\phi$. For further discussion, see also~\cite{Lugo:1994ra}. As it is expected, the central charge of this CFT matches that of the IR fixed point in the original theory. Indeed, the central charge now decomposes as
\begin{equation}
\label{c.split}
c_\text{IR} = c_{su(2)/u(1)} + c_{\text{l.d.}} + 1 = 4 + \frac{18}{k} + \mathcal{O}\left( \frac{1}{k^2} \right)\,,
\end{equation}
where $c_\text{$\ell$.d}=1+6s\alpha'Q^2$ is the central charge of the spacelike linear dilaton, with $s=1$, $\alpha'=\nicefrac1k$, and background charge $Q=-2$ (see also the discussion in Section~2.2 of~\cite{Itsios:2021wso}). The above value coincides with the value of the $c$-function \eqref{Cfunction.noncompact} for $\l = 0$.

\noindent \textbf{The $\l \to 1$ limit:} In this case the corresponding target space metric \eqref{LambdaDeformedMetric} blows up. Nevertheless, the singular behaviour of the limit can be avoided by setting
\begin{equation}
\label{asymptotic.limit.one}
 \l = 1 - \frac{\kappa^2}{k} \, , \qquad
 \th = - \frac{\kappa^2}{k} z_1 \, , \qquad
 \psi = \pi + \frac{\kappa^2}{k} (z_2 - 2 z_3) \, , \qquad
 \phi = \frac{\kappa^2}{k} z_3 
\end{equation}
and then we send $k$ to infinity. As a result, the line element of the target space metric reads
\begin{eqnarray}
 \label{Lambda1Metric}
  ds^2&=& \kappa^2\Bigg(\big( 1 + z^2_1 + z^2_2 \big) d\tau^2 
 + 4 \, \big( z_1 \, dz_1 + z_2 \, dz_2 - z_2 \, dz_3 \big) d\tau
 + 4 \, dz^2_1
 + \frac{1 + z^2_1 + z^2_2}{z^2_1} dz^2_2 \nonumber
 \\[5pt]
 && + 4 \, \frac{1 + z^2_2}{z^2_1} \, dz^2_3
 + 4 \, \frac{z_2}{z_1} dz_1 \, dz_2
 - 8 \, \frac{z_2}{z_1} dz_1 \, dz_3
 - 4 \, \frac{1 + z^2_2}{z^2_1} dz_2 \, dz_3\Bigg) \, .
\end{eqnarray}
This geometry exhibits a curvature singularity at $z_1 = 0$ where $R = - \frac{1 + 3 z^2_1 + z^2_2}{\kappa^2z^2_1}$. Also, $\partial_\tau$ and $\partial_{z_3}$ are Killing vectors reflecting the invariance of the metric under shifts of $\tau$ and $z_3$ respectively.  The leading behaviour for the scalar is instead
\begin{equation}
 \label{Lambda1AsymptGeom}
 e^{\, - 2 \Phi} = \frac{3}{4} e^{\, 4 \tau} z^2_1 \,,
\end{equation}
which clearly diverges when $z_1 = 0$. It is worth noting that the background~\eqref{Lambda1Metric} does not arise as an asymptotic limit of the geometric $SU(2,1)/U(2)$ coset, in contrast to the cases analyzed in~\cite{Itsios:2021wso} for the deformed $SO(d,1)_{-k}/SO(d)_{-k}$ coset CFTs. The precise identification of the $\s$-model whose target space is described by \eqref{Lambda1Metric} remains an open problem.

The RG flow of the parameter $\kappa^2$ can be found using \eqref{RGequation}
\begin{equation}
\frac{d\kappa^2}{dt}=-\frac32\,,\quad \xi=G^{\mu\nu}\partial_\nu\Phi\partial_\mu
-\frac{3}{2\kappa^2}\left(z_1\partial_{z_1}+z_2\partial_{z_2}+z_3\partial_{z_3}\right)\,,
\end{equation}
which is consistent with \eqref{RGs.oneloop.su12ovu2} in the limit \eqref{asymptotic.limit.one}.


\section{A scale invariant $\s$-model}
\label{ScaleInvariantModel}

Another interesting $\s$-model can be derived from the one discussed in Section \ref{TheNCLambdaModel} by making the coordinate $\phi$ non-compact. This is realised by applying the following analytic continuation
\begin{equation}
 \label{AnalyticContinuationScaleInvariance}
 \phi \to \pi + i \, \frac{2}{3} \, \ln u \, , \qquad x_0 \to i x_0 \, ,
\end{equation}
where the last ensures that the $\s$-model is real. After such a transformation, the asymptotic limit where the coordinate $u$ becomes large makes sense provided that
$\l$ becomes small such that $\l u$ is kept fixed. Indeed, if we apply the aforementioned transformations in \eqref{LambdaDeformedMetric}, \eqref{LambdaZeroMetricNC}, \eqref{MetricDefNC} the resulting metric is now linear in $\l$. In particular it has the form
\begin{equation}
 \label{MetricScale}
 G_{\m\n} = 2 k \big( \bg_{\m\n} + \l g_{\m\n} \big) \,,
\end{equation}
and exhibits two time-like directions.
The components of $\bg_{\m\n}$ now are
\begin{equation}
 \label{MetricScaleZerolambda}
 \begin{aligned}
  & \bg_{uu} = - \frac{4 (1 + v^2)^2 + (1 + v)^2 (3 - 2 v + 3 v^2) x^2_0 - (1 - v)^2 (3 + 2 v + 3 v^2) x^2_1}{9 u^2 (1 - v^2)^2 \tilde{\r}^2} \, ,
  \\[5pt]
  & \bg_{vv} = \frac{1}{v^2} \, , \qquad
  \bg_{x_0 x_0} = - \Big( \frac{1 + v}{1 - v} \Big)^2 \frac{1}{\tilde{\r}^2} \, , \qquad
  \bg_{x_1 x_1} = \Big( \frac{1 - v}{1 + v} \Big)^2 \frac{1}{\tilde{\r}^2} \, ,
  \\[5pt]
  & \bg_{u x_0} = - \frac{1}{3 u} \Big( \frac{1 + v}{1 - v} \Big) ^2\frac{x_1}{\tilde{\r}^2} \, , \qquad
  \bg_{u x_1} = \frac{1}{3 u} \Big( \frac{1 - v}{1 + v} \Big) ^2\frac{x_0}{\tilde{\r}^2} \, ,
 \end{aligned}
\end{equation}
while those for $g_{\m\n}$
\begin{equation}
 \label{MetricScaleNonZerolambda}
 \begin{aligned}
  & g_{uu} = - \frac{x_0 + x_1}{9 u (1 - v^2)^2 \tilde{\r}^2} \left( 8 v (1 + v^2) + \Big( (1 + v)^2 x_0 - (1 - v)^2 x_1 \Big)^2 \right) \, ,
  \\[5pt]
  & g_{vv} = \frac{u(x_1 - x_0)}{v^2} \, , \qquad
  g_{x_0 x_0} = - u \Big( \frac{1 + v}{1 - v} \Big)^2 \frac{x_0 + x_1}{\tilde{\r}^2} \, ,
  \\[5pt]
  & g_{x_1 x_1} = - u \Big( \frac{1 - v}{1 + v} \Big)^2 \frac{x_0 + x_1}{\tilde{\r}^2} \, , \qquad
  g_{uv} = \frac{1}{3 v} \frac{(1 + v)^2 x_0 - (1 - v)^2 x_1}{1 - v^2} \, ,
  \\[5pt]
  & g_{u x_0} = - \frac{1}{3} \frac{(1 + v^2) (1 + \tilde{\r}^2) + 2 v (x_0 + x_1)^2}{(1 - v)^2\tilde{\r}^2} \, ,
  \\[5pt]
  & g_{u x_1} = \frac{1}{3} \frac{(1 + v^2) (1 + \tilde{\r}^2) + 2 v (x_0 + x_1)^2}{(1 + v)^2\tilde{\r}^2} \, , \qquad
  g_{v x_0} = \frac{u}{v} \frac{1 + v}{1 - v} \, ,
  \\[5pt]
  & g_{v x_1} = - \frac{u}{v} \frac{1 - v}{1 + v} \, , \qquad
  g_{x_0 x_1} = u \frac{x_0 + x_1}{\tilde{\r}^2} \, .
 \end{aligned}
\end{equation}
In the above expressions we defined for convenience the combinations
\begin{equation}
 \tau = \ln v \, , \qquad \tilde{\r} = \sqrt{1 + x^2_0 - x^2_1} \, .
\end{equation}

In the same way, the analytic continuation \eqref{AnalyticContinuationScaleInvariance} together with the asymptotic limit $u \gg 1$ with $\l u$ fixed implies the following expression for the scalar $\Phi$ of \eqref{LambdaScalarNC}
\begin{equation}
 \label{ScalarScale}
 e^{- 2 \Phi} = \frac{3 k^8}{16 \l^4} \Big( \frac{1 - v^2}{v} \Big)^4 \tilde{\r}^2 \, .
\end{equation}
Notice that this is the leading term in the expansion for small $\l$.

One can also verify that the metric \eqref{MetricScale}, \eqref{MetricScaleZerolambda}, \eqref{MetricScaleNonZerolambda} and the scalar \eqref{ScalarScale} imply
\begin{equation}
 \begin{aligned}
  & \b_{\Phi} = R + 4 \, \nabla^2 \Phi - 4 \, \big( \partial \Phi \big)^2 = - \frac{6}{k} \, ,
  \\[5pt]
  & \b_{G_{\m\n}} = R_{\m\n} + 2 \nabla_\m \partial_\n \Phi = 3 g_{\m\n} \, .
 \end{aligned}
\end{equation}

\subsubsection*{Scale invariance}

The renormalisability of this $\s$-model at one loop is guaranteed provided that $\l$ runs according to
\begin{equation}
 \label{BetalambdaScale}
 \frac{d \l}{d t} = \frac{3 + 2 c}{2 k} \l \, .
\end{equation}
The above condition is implied from the RG flow equation \eqref{RGequation}, where the vector $\xi$ is given by
\footnote{The vector $\eta$ satisfies the relation
$
\nabla_\mu\eta_\nu+\nabla_\nu\eta_\mu=\frac{c}{k}g_{\mu\nu}\,,
$
and thus generates the perturbation. 
}
\begin{equation}
 \xi = \partial^\m \Phi \, \partial_{\m} + \eta\,,\quad \eta=\frac{c}{k} \, u \, \partial_u\,,  \quad c = \text{const} \, .
\end{equation}
Notice that the parameter $\l$ can be absorbed in the metric by redefining the coordinate $u$ as $u \to \frac{u}{\l}$. This case is reminiscent of the class of examples constructed in \cite{Itsios:2021eig}. In particular, for $\lambda\neq0$ one can take $\l = 1$. Going to \eqref{BetalambdaScale}, this means that the constant $c$ must be fixed to the non-zero value $c = - \frac{3}{2}$ corresponding to a scale rather than a Weyl invariant model. This follows from the fact that the components $\xi_{\m}$ cannot, for $c\neq0$, be written as derivatives of a scalar. Consequently, the $\s$-model discussed here is scale invariant but not conformal invariant~\cite{Hull:1985rc}. 

Finally, a brief remark is in order regarding the scale-invariant model presented here, in contrast to those discussed in~\cite{Itsios:2021eig}. The model considered in this work, which is based on a unitary group, does not admit a deformation in the form of a Kerr-Schild perturbation of the underlying coset CFT. This is in contrast to the models in~\cite{Itsios:2021eig}, which are based on orthogonal groups and do allow for such a deformation
\footnote{In the present case, the deformation is realised by the metric $g_{\mu\nu}$ as given in~\eqref{MetricScale}, which is a rank-2 tensor and therefore not of the Kerr-Schild form, i.e. $g_{\mu\nu}\sim\ell_\mu\ell_\nu$, where $\ell_\mu$ is a null and geodesic vector.}.

\subsubsection*{Restoring conformal invariance}

Conformal invariance at one-loop order can be restored for large values of $x_0 , \, x_1$ and $v$ while keeping $x^2_0 u$, $x^2_1 u$ and $v^2 u$ fixed. Indeed, in this limit the metric \eqref{MetricScale}, \eqref{MetricScaleZerolambda}, \eqref{MetricScaleNonZerolambda} becomes
\begin{equation}
 \label{MetricRestoredConformalInvariance}
 ds^2 = 2 k \Bigg( \frac{dv^2}{v^2} - \frac{du^2}{3 u^2} - \frac{dx^2_0 - dx^2_1}{x^2_0 - x^2_1} - \frac{2}{3} \frac{x_1 dx_0 - x_0 dx_1}{x^2_0 - x^2_1} \frac{du}{u} \Bigg) 
\end{equation}
and the $\l$-dependence drops out. In addition, the scalar \eqref{ScalarScale} simplifies to
\be
\label{ScalarScale1}
e^{-2\Phi}=\frac{3}{16}  k^8 v^4 \left(x_0^2-x_1^2\right)\,,
\ee
after neglecting an additional infinite constant.

It turns out that the above line element describes a flat geometry. This can be seen if we change coordinates to
\begin{equation}
 \begin{aligned}
  & x_0 = e^{y_0} \sinh \Big( \frac{\sqrt{3}}{2} y_2 \Big) \, , \qquad x_1 = e^{y_0} \cosh \Big( \frac{\sqrt{3}}{2} y_2 \Big) \, ,
  \\[5pt]
  & v = e^{y_3} \, , \qquad u = e^{\sqrt{3} \big( y_1 + \frac{y_2}{2} \big)} \, .
 \end{aligned}
\end{equation}
In these coordinates the metric \eqref{MetricRestoredConformalInvariance} reads
\begin{equation}
 ds^2 = 2 k \big( - dy^2_0 - dy^2_1 + dy^2_2 + dy^2_3 \big)\,,
\end{equation}
describing the space $\mathbb{R}^{2,2}$. Correspondingly, the scalar $\Phi$ in \eqref{ScalarScale1} becomes linear in $y_0$ and $y_3$
\begin{equation}
 \Phi = - y_0 - 2 y_3\,,
\end{equation}
where we dropped a constant term
\footnote{The central charge of the aforementioned solution is still given by $c=4+\nicefrac{18}{k}$, as derived using either $\b_\Phi$ in Eq. \eqref{betaPhibetaGNC} or the linear dilaton backgrounds (timelike $y_0$ and spacelike $y_3$), following the discussion after Eq.\eqref{c.split}.}.


\section{Supergravity embeddings}
\label{Sec:SUGRA}

In this Section, we promote the $\l = 0$ and $\l \to 1$ limits of the asymptotic model discussed in Section \ref{AsymptoticModel} into solutions of type-II supergravity
\footnote{We will not consider supergravity embedding of the scale invariant action studied in Section~\ref{ScaleInvariantModel}, due to the two timelike directions of \eqref{MetricScale}.}.


\subsection{Type-II embeddings of the $\l = 0$ model}

Below we provide three solutions of the type-II supergravity which accommodate the field content of the asymptotic model for $\l = 0$. The first solution has trivial RR sector while the other two are backgrounds of the type-IIA supergravity.

\noindent\textbf{Type-II solution with trivial RR sector:} The ten-dimensional geometry of this background is of the form $AdS_3 \times S^3 \times \cM_4$, where $\cM_4$ is the space with metric \eqref{Lambda0AsymptGeom}. We also choose the normalisation of $AdS_3$ and $S^3$ to be $R_{\m\n} = - \frac{2}{L^2_1} g_{\m\n}$ and $R_{\m\n} = \frac{2}{L^2_2} g_{\m\n}$ respectively. The ten-dimensional geometry is also supported by the dilaton \eqref{Lambda0AsymptScalar} and the NS three-form
\begin{equation}
 H_3 = \frac{2}{L_1} \, \text{Vol}(AdS_3) + \frac{2}{L_2} \, \text{Vol}(S^3) \, .
\end{equation}
It is clear that $\text{Vol}(AdS_3)$ and $\text{Vol}(S^3)$ refer to the volume forms on $AdS_3$ and $S^3$ respectively. To ensure that the above content solves the equations of motion of the type-II supergravity, the radii $L_1$ and $L_2$ must satisfy the following condition
\footnote{The condition in \eqref{radii.condition} can be waived by modifying the $\tau$-dependence of the dilaton in \eqref{Lambda0AsymptScalar} as 
$$
e^{-2\Phi}=\frac{3}{16} k^8 \, e^{-2\ell \tau} \sin^2\th\,,\quad \ell=\sqrt{1-\frac{2k}{L_1^2}+\frac{2k}{L_2^2}}\,.
$$
We thank A. Tseytlin for raising this point on the $AdS_3\times S^3\times S^3\times S^1$.}
\begin{equation}
\label{radii.condition}
 \frac{1}{L^2_2} = \frac{1}{L^2_1} + \frac{3}{2 k} \, .
\end{equation}
Notice that one can also derive a type-IIB solution via S-duality.

\noindent\textbf{Type-IIA solution (i):} The ten-dimensional background now has a geometry whose form is $AdS_4 \times S^2 \times \cM_4$ with $\cM_4$ being again the space with metric \eqref{Lambda0AsymptGeom}. The normalisation for the $AdS_4$ and $S^2$ factors is chosen to be $R_{\m\n} = - \frac{3}{L^2_1} g_{\m\n}$ and $R_{\m\n} = \frac{1}{L^2_2} g_{\m\n}$ respectively. This spacetime is supported by the dilaton
\begin{equation}
 \label{L0DilatonAdS4S2}
 e^{- 2 \Phi} = \frac{3}{16} k^8 \, e^{\tau} \sin^2\th \,,
\end{equation}
Notice that the part that is linear in $\tau$ differs by an overall constant with respect to the one in \eqref{Lambda0AsymptScalar}. Among the form fields, only the RR ones are non-trivial and their expressions are given below
\begin{equation}
\label{L0DilatonAdS4S2.RR}
 \begin{aligned}
  & F_2 = s_1 \, \frac{\sqrt{3 L^2_2 - L^2_1}}{L_1 L_2} \, e^{- \Phi} \big( e^1 \wedge e^3 - e^2 \wedge e^4 \big) \, , \quad
  \\[5pt]
  & F_4 = s_2 \, \frac{\sqrt{3 L^2_2 + L^2_1}}{L_1 L_2} \, e^{- \Phi} \big( e^1 \wedge e^4 + e^2 \wedge e^3 \big) \wedge \text{Vol}(S^2) \, ,
 \end{aligned}
\end{equation}
where $\text{Vol}(S^2)$ is understood to be the volume form on the two-sphere and $s_{1,2} = \pm 1$. To present the RR forms we chose the following unusual orthogonal frame on $\cM_4$
\begin{equation}
 \label{L0frame}
 \begin{aligned}
  & e^1 = \sqrt{2 k} \, d\tau \, , \quad\quad 
  e^2 = - \sqrt{2 k} \, d\phi \, ,
  \\[5pt]
  & e^3 = - \sqrt{2 k} \, \left( \cos \Big( \psi + \frac{\phi}{2} \Big) \, d\th + \cot\th \, \sin \Big( \psi + \frac{\phi}{2} \Big) \, d\psi \right) \, ,
  \\[5pt]
  & e^4 = \sqrt{2 k} \, \left( \sin \Big( \psi + \frac{\phi}{2} \Big) \, d\th - \cot\th \, \cos \Big( \psi + \frac{\phi}{2} \Big) \, d\psi \right) \, .
 \end{aligned}
\end{equation}

The solution of the type-IIA equations of motion is guaranteed if
\begin{equation}
 \frac{6}{L^2_1} =   \frac{1}{L^2_2} + \frac{3}{4 k} \, .
\end{equation}
Notice that in order for the solution to be real, the radii must be further restricted as $L_1 \leqslant \sqrt{3} L_2$. This in conjunction with the previous condition implies that $L_1$ must be bounded as $2 \sqrt{k} \leqslant L_1 \leqslant 2 \sqrt{3 k}$.

The background given by \eqref{L0DilatonAdS4S2}, \eqref{L0DilatonAdS4S2.RR}, and \eqref{L0frame} reduces to one with geometry $AdS_4\times T^2\times \cM_4$ in the limit $L_2 \to \infty$, accompanied by an appropriate zoom-in on the two-sphere.

\noindent\textbf{Type-IIA solution (ii):} A second type-IIA background with geometry of the form $AdS_2 \times S^4 \times \cM_4$ can also be constructed. This is achieved via a double analytic continuation in the previous example which transforms $AdS_4$ to an $S^4$ and $S^2$ to an $AdS_2$. Here the $AdS_2$ and $S^4$ spaces are normalised such that $R_{\m\n} = - \frac{1}{L^2_1} g_{\m\n}$ and $R_{\m\n} = \frac{3}{L^2_2} g_{\m\n}$ respectively. The geometry is supported by the dilaton \eqref{L0DilatonAdS4S2} and a set of RR fluxes which in this case read
\begin{equation}
\label{RR24.IIA}
 \begin{aligned}
  & F_2 = s_1 \, \frac{\sqrt{L^2_2 - 3 L^2_1}}{L_1 L_2} \, e^{- \Phi} \big( e^1 \wedge e^3 - e^2 \wedge e^4 \big) \, , \qquad
  \\[5pt]
  & F_4 = s_2 \, \frac{\sqrt{3 L^2_1 + L^2_2}}{L_1 L_2} \, e^{- \Phi} \big( e^1 \wedge e^4 + e^2 \wedge e^3 \big) \wedge \text{Vol}(AdS_2) \, .
 \end{aligned}
\end{equation}
Now $\text{Vol}(AdS_2)$ is understood as the volume form on $AdS_2$, $e^a \, (a = 1,2,3,4)$ refers to the frame \eqref{L0frame} and $s_{1,2} = \pm 1$. To ensure that the type-IIA equations of motion are solved, one has also to impose the condition
\begin{equation}
\label{RR24.IIA.cont}
  \frac{1}{L^2_1} =  \frac{6}{L^2_2}+ \frac{3}{4 k} \, .
\end{equation}
The RR forms are now real provided that $L_2 \geqslant \sqrt{3} L_1$. When the latter is combined with the condition above it gives $L_1 < 2 \sqrt{\nicefrac{k}{3}}$.

Finally, a background with geometry $AdS_2 \times T^4 \times \cM_4$ can be obtained from the one above by taking the limit $L_2 \to \infty$, combined with an appropriate zoom-in on the four-sphere.

\subsection{A type-IIA embedding for the $\l \to 1$ limit}

Unlike the $\l = 0$ case, we were able to find only one embedding of the $\l \to 1$ model in type-II supergravity. The corresponding background has a geometry of the form $AdS_2 \times S^4 \times \cM_4$, where now $\cM_4$ is the space with metric \eqref{Lambda1Metric}. Again $AdS_2$ and $S^4$ are chosen such that $R_{\m\n} = - \frac{1}{L^2_1} g_{\m\n}$ and $R_{\m\n} = \frac{3}{L^2_2} g_{\m\n}$ respectively. Besides the metric, the NS sector of this background contains the dilaton \eqref{Lambda1AsymptGeom}. On the other hand, the RR sector consists of the following fields
\begin{equation}
 \begin{aligned}
  & F_2 = s_1 \, \frac{\sqrt{L^2_2 - 3 L^2_1}}{L_1 L_2} \, e^{- \Phi} \big( e^1 \wedge e^3 - e^2 \wedge e^4 \big) \, , \qquad
  \\[5pt]
  & F_4 = \frac{3}{L_2} \, e^{- \Phi} \Bigg( s_2 \, e^1 \wedge e^4
  + s_3 \, \sqrt{\frac{1 - 2 \, L^2_1}{1 + 6 \, L^2_1}} \, e^2 \wedge e^3 \Bigg) \wedge \text{Vol}(AdS_2) \, ,
 \end{aligned}
\end{equation}
where again $\text{Vol}(AdS_2)$ refers to the volume form on $AdS_2$ and $s_{1,2,3} = \pm 1$. Moreover, we defined the following orthogonal frame on $\cM_4$
\begin{equation}
 \begin{aligned}
  & e^1 = \frac{d\tau}{\sqrt{2}} \, , \qquad\qquad 
     e^2 = - \frac{dz_2 + z_2 \, d\tau}{\sqrt{2}} \, ,
 \\[5pt]
  & e^3 = \frac{z_1}{\sqrt{2}} \, d\tau + \sqrt{2} \, dz_1 + \frac{1}{\sqrt{2}} \frac{z_2}{z_1} \, dz_2 - \sqrt{2} \frac{z_2}{z_1} \, dz_3 \, , \qquad
  e^4 = \frac{dz_2 - 2 \, dz_3}{\sqrt{2} \, z_1} \, .
 \end{aligned}
\end{equation}
The above content is a solution of the type-IIA supergravity provided that the radii $L_1$ and $L_2$ satisfy the condition
\begin{equation}
 \frac{6}{L^2_2} = 6 + \frac{1}{L^2_1} \, .
\end{equation}
To ensure that the RR fluxes are not imaginary we should also take into account the restrictions $L_2 \geqslant \sqrt{3} L_1$ and $L_1 \leqslant \nicefrac{1}{\sqrt{2}}$. All together, the aforementioned conditions imply that actually $L_1 \leqslant \nicefrac{1}{\sqrt{6}}$.

\section{Conclusion and Outlook}

The main objective of this work is to explore deformations of coset CFTs with non-compact target spaces, focusing on models based on unitary groups. This extends previous results obtained for orthogonal groups \cite{Itsios:2021wso,Itsios:2021eig}. We begin by studying a deformation of a coset CFT realised in terms of a gauged WZW model on the symmetric $SU(3)_k /U(2)_k$ space. The deformed model is classically integrable \cite{Hollowood:2014rla,Hollowood:2015dpa} and renormalisable at one-loop order \cite{Sfetsos:2014jfa,Appadu:2015nfa}. We then turn to its non-compact analogue corresponding to the coset $SU(2,1)_{-k}/U(2)_{-k}$. This theory exhibits an interesting asymptotic limit when one of the fields becomes large, reducing to an integrable deformation of the IR $SU(2)_k/U(1)$ coset CFT, supplemented by a linear dilaton and a free boson \cite{Lugo:1994ra}. In the UV, however, the model does not approach the geometric $SU(2,1)/U(2)$ coset, in contrast to the cases considered in \cite{Itsios:2021wso}.

By considering an alternative asymptotic and double-scaling limit of the deformed $SU(2,1)_{-k}/U(2)_{-k}$ model, we derived a two-dimensional field theory that is scale-invariant but not conformally invariant at one-loop order. Notably, this deformation is not of the Kerr-Schild type, in contrast to the examples studied in \cite{Itsios:2021eig}. In a subsequent asymptotic limit, Weyl invariance is restored resulting to a theory with two free bosons and two linear dilaton fields.

A natural continuation of the present work would be to extend the analysis to higher-dimensional $\l$-deformed symmetric coset CFTs based on unitary groups, with the aim of identifying scale-invariant models. Additionally, scale-invariant deformations could be explored in asymptotic double-scaling limits of the Yang-Baxter deformed symmetric cosets constructed in \cite{Delduc:2013fga,Delduc:2013qra}. These models are related to the $\l$-deformed ones via Poisson-Lie T-duality combined with an analytic continuation \cite{Vicedo:2015pna, Hoare:2015gda, Sfetsos:2015nya, Klimcik:2015gba, Klimcik:2016rov}.

A more challenging question we aim to address in future work is whether the asymptotic model constructed in Sec. \ref{AsymptoticModel} can be embedded into type-II supergravity or generalised supergravity~\cite{Arutyunov:2015mqj}, capturing all allowed values of $\l$ in the interval $[0 , 1)$. An even more ambitious direction would be to promote the $\l$-models on $SU(3)_k/U(2)_k$ and $SU(2,1)_{-k}/U(2)_{-k}$ to full solutions of type-II supergravity or generalised supergravity~\cite{Arutyunov:2015mqj}, analogous to the constructions achieved in \cite{Itsios:2024gqr} for deformed coset CFTs based on orthogonal groups.

\section*{Acknowledgements}

We would like to thank Y.~ Nakayama, K.~Sfetsos and A.~Tseytlin for useful correspondence.
The research work of G. Itsios is supported by the Einstein Stiftung Berlin via the Einstein International Postdoctoral Fellowship program ``Generalised dualities and their holographic applications to condensed matter physics'' (project number IPF-2020-604). K. Siampos would like to thank the theory division at CERN for hospitality and financial support at the last stages of the present work.

\appendix


\section{The Gell-Mann matrices}
\label{GellMannMatrices}

Here we collect the Gell-Mann matrices which provide a basis for the Lie algebra of the $SU(3)$ group and its non-compact form $SU(2,1)$. These are
\begin{equation}
 \begin{aligned}
  \n_1 & =
  \begin{pmatrix}
   0 & 1 & 0
   \\
   1 & 0 & 0
   \\
   0 & 0 & 0
  \end{pmatrix} \, , \qquad\quad
  \n_2 =
  \begin{pmatrix}
   0 & -i & 0
   \\
   i & 0 & 0
   \\
   0 & 0 & 0
  \end{pmatrix} \, , \qquad
  \n_3 =
  \begin{pmatrix}
   1 & 0 & 0
   \\
   0 & -1 & 0
   \\
   0 & 0 & 0
  \end{pmatrix} \, , \qquad
  \\[5pt]
  \n_4 & =
  \begin{pmatrix}
   0 & 0 & 1
   \\
   0 & 0 & 0
   \\
   1 & 0 & 0
  \end{pmatrix} \, , \qquad\quad
  \n_5 =
  \begin{pmatrix}
   0 & 0 & -i
   \\
   0 & 0 & 0
   \\
   i & 0 & 0
  \end{pmatrix} \, , \qquad
  \n_6 =
  \begin{pmatrix}
   0 & 0 & 0
   \\
   0 & 0 & 1
   \\
   0 & 1 & 0
  \end{pmatrix} \, , \qquad
 \\[5pt]
 \n_7 & =
 \begin{pmatrix}
   0 & 0 & 0
   \\
   0 & 0 & -i
   \\
   0 & i & 0
  \end{pmatrix} \, , \qquad
 \n_8 = \frac{1}{\sqrt{3}}
  \begin{pmatrix}
   1 & 0 & 0
   \\
   0 & 1 & 0
   \\
   0 & 0 & -2
  \end{pmatrix} \, .
 \end{aligned}
\end{equation}
It is easy to check that $\n_1$, $\n_2$ and $\n_3$ close into an $SU(2)$ algebra and that $\n_8$ commutes with all $\n_i$, where $i = 1, 2, 3$. Therefore, the group $SU(3)$ admits a $U(2) = SU(2) \times U(1)$ subgroup, generated by $\n_1, \, \n_2, \, \n_3$ and $\n_8$. The above matrices are all traceless and Hermitian and also satisfy the following relation
\begin{equation}
 \tr \big( \n_A \n_B \big) = 2 \, \d_{AB} \, , \qquad A, B = 1, \ldots ,8 \, .
\end{equation}



\begin{thebibliography}{99}

\bibitem{Wess:1960}
J.~Wess, {\it The conformal invariance in quantum field theory}, 
\href{https://link.springer.com/article/10.1007/BF02733168}{Nuovo Cimento {\bf 18} (1960), 1086}.

\bibitem{Gross:1970tb}
D.~J.~Gross and J.~Wess,
{\it Scale invariance, conformal invariance, and the high-energy behavior of scattering amplitudes},
\href{https://journals.aps.org/prd/abstract/10.1103/PhysRevD.2.753}{Phys. Rev. \textbf{D2} (1970), 753}.

\bibitem{El-Showk:2011xbs}
S.~El-Showk, Y.~Nakayama and S.~Rychkov,
{\it What Maxwell Theory in D{\ensuremath{<}}{\ensuremath{>}}4 teaches us about scale and conformal invariance},
\href{https://www.sciencedirect.com/science/article/abs/pii/S0550321311001465?via\%3Dihub}{Nucl. Phys. \textbf{B848} (2011), 578},
\href{https://arxiv.org/abs/1101.5385}{\tt arXiv:1101.5385 [hep-th]}.

\bibitem{Gimenez-Grau:2023lpz}
A.~Gimenez-Grau, Y.~Nakayama and S.~Rychkov,
{\it Scale without conformal invariance in dipolar ferromagnets},
\href{https://journals.aps.org/prb/abstract/10.1103/PhysRevB.110.024421}{Phys. Rev. \textbf{B110} (2024) no.2, 024421},
\href{https://arxiv.org/abs/2309.02514}{\tt arXiv:2309.02514 [hep-th]}.

\bibitem{Papadopoulos:2024uvi}
G.~Papadopoulos and E.~Witten,
{\it Scale and conformal invariance in 2d {\ensuremath{\sigma}}-models, with an application to $\mathcal{N}$ = 4 supersymmetry},
\href{https://link.springer.com/article/10.1007/JHEP03(2025)056}{JHEP \textbf{03} (2025), 056},
\href{https://arxiv.org/abs/2404.19526}{\tt arXiv:2404.19526 [hep-th]}.

\bibitem{Chawla:2025fpn}
L.~Chawla and M.~Flory,
{\it Scale without Conformal Invariance in bottom-up Holography},
\href{https://arxiv.org/abs/2505.09692}{\tt arXiv:2505.09692 [hep-th]}.

\bibitem{Nakayama:2024jwq}
Y.~Nakayama,
{\it Parisi-Sourlas Supertranslation and Scale Invariance without Conformal Symmetry},
\href{https://journals.aps.org/prl/abstract/10.1103/PhysRevLett.134.131602}{Phys. Rev. Lett. \textbf{134} (2025) no.13, 131602},
\href{https://arxiv.org/abs/2411.12934}{\tt arXiv:2411.12934 [hep-th]}.

\bibitem{Nakayama:2013is}
Y.~Nakayama,
{\it Scale invariance vs conformal invariance},
\href{https://www.sciencedirect.com/science/article/pii/S0370157314004499?via\%3Dihub}{Phys. Rept. \textbf{569} (2015), 1},
\href{https://arxiv.org/abs/1302.0884}{\tt arXiv:1302.0884 [hep-th]}.

 \bibitem{Polchinski.Scale}
J. Polchinski, {\it Scale and conformal invariance in quantum field theory},
\href{https://www.sciencedirect.com/science/article/pii/0550321388901794}{Nucl.Phys. {\bf B303} (1988), 226}.

\bibitem{Zamolodchikov:1986gt}
A.~B.~Zamolodchikov,
{\it Irreversibility of the Flux of the Renormalization Group in a 2D Field Theory},
\href{http://jetpletters.ru/ps/1413/article_21504.shtml}{JETP Lett. \textbf{43} (1986), 730}.

 \bibitem{Lucher.Mack}
M. L\"uscher, G. Mack, {\it The energy momentum tensor of critical quantum field
theories in 1+1 dimensions}, unpublished (1976).\\ 
G. Mack, {\it Introduction to conformal invariant quantum
field theory in two or more dimensions}, in Nonperturbative Quantum Field Theory.
Proceedings,  \href{https://link.springer.com/chapter/10.1007/978-1-4613-0729-7_12}{Nato Science Series B: Physics Vol. 185, pp 353}.

\bibitem{Hull:1985rc}
C.~M.~Hull and P.~K.~Townsend,
{\it Finiteness and Conformal Invariance in Nonlinear $\sigma$-Models},
\href{https://www.sciencedirect.com/science/article/pii/0550321386902890?via\%3Dihub}{Nucl. Phys. \textbf{B274} (1986), 349}.

\bibitem{Itsios:2021eig}
G.~Itsios, K.~Sfetsos and K.~Siampos,
{\it Kerr\textendash{}Schild perturbations of coset CFTs as scale invariant integrable \ensuremath{\sigma}-models},
\href{https://www.sciencedirect.com/science/article/pii/S0550321321002911?via\%3Dihub}{Nucl. Phys. \textbf{B973} (2021), 115594},
\href{https://arxiv.org/abs/2109.05040}{\tt arXiv:2109.05040 [hep-th]}.

\bibitem{Sfetsos:2013wia}
K.~Sfetsos,
{\it Integrable interpolations: From exact CFTs to non-Abelian T-duals},
\href{https://www.sciencedirect.com/science/article/pii/S0550321314000066?via\%3Dihub}{Nucl. Phys. \textbf{B880} (2014), 225},
\href{https://arxiv.org/abs/1312.4560}{{\tt arXiv:1312.4560 [hep-th]}}.

\bibitem{Hollowood:2014rla}
T.~J.~Hollowood, J.~L.~Miramontes and D.~M.~Schmidtt,
{\it Integrable Deformations of Strings on Symmetric Spaces},
\href{https://link.springer.com/article/10.1007/JHEP11(2014)009}{JHEP \textbf{11} (2014), 009},
\href{https://arxiv.org/abs/1407.2840}{\tt arXiv:1407.2840 [hep-th]}.

\bibitem{Hollowood:2015dpa}
T.~J.~Hollowood, J.~L.~Miramontes and D.~M.~Schmidtt,
{\it S-Matrices and Quantum Group Symmetry of k-Deformed Sigma Models},
\href{https://iopscience.iop.org/article/10.1088/1751-8113/49/46/465201}{J. Phys. \textbf{A49} (2016) no.46, 465201},
\href{https://arxiv.org/abs/1506.06601}{\tt arXiv:1506.06601 [hep-th]}.

\bibitem{Sfetsos:2014jfa}
K.~Sfetsos and K.~Siampos,
{\it Gauged WZW-type theories and the all-loop anisotropic non-Abelian Thirring model},
\href{https://www.sciencedirect.com/science/article/pii/S0550321314001953?via\%3Dihub}{Nucl. Phys. \textbf{B885} (2014), 583},
\href{https://arxiv.org/abs/1405.7803}{\tt arXiv:1405.7803 [hep-th]}.

\bibitem{Appadu:2015nfa}
C.~Appadu and T.~J.~Hollowood,
{\it Beta function of k deformed AdS$_{5}$ \texttimes{} S$^{5}$ string theory},
\href{https://link.springer.com/article/10.1007/JHEP11(2015)095}{JHEP \textbf{11} (2015), 095},
\href{https://arxiv.org/abs/1507.05420}{\tt arXiv:1507.05420 [hep-th]}.

\bibitem{Eichenherr:1979ci}
H.~Eichenherr and M.~Forger,
{\it On the Dual Symmetry of the Nonlinear Sigma Models},
\href{https://www.sciencedirect.com/science/article/abs/pii/0550321379902761?via\%3Dihub}{Nucl. Phys. \textbf{B155} (1979), 381-393}.

\bibitem{Curtright:1994be}
  T.~Curtright and C.~K.~Zachos,
  {\it Currents, charges, and canonical structure of pseudodual chiral models},
  \href{https://journals.aps.org/prd/abstract/10.1103/PhysRevD.49.5408}{Phys. Rev. {\bf D49} (1994) 5408},
  \href{http://arxiv.org/abs/hep-th/9401006}{\tt hep-th/9401006}.

\bibitem{Bykov:2014efa}
D.~Bykov,
{\it Integrable properties of sigma-models with non-symmetric target spaces},
\href{https://www.sciencedirect.com/science/article/pii/S0550321315000875?via\%3Dihub}{Nucl. Phys. \textbf{B894} (2015), 254-267},
\href{https://arxiv.org/abs/1412.3746}{\tt arXiv:1412.3746 [hep-th]}.

\bibitem{Lugo:1994ra}
A.~R.~Lugo,
{\it Gravitational instantons and black plane solutions in 4-D string theory},
\href{https://journals.aps.org/prd/abstract/10.1103/PhysRevD.52.2266}{Phys. Rev. \textbf{D52} (1995), 2266},
\href{https://arxiv.org/abs/hep-th/9411152}{{\tt hep-th/9411152}}.

\bibitem{Maillet2}
  J.~M.~Maillet,
 {\it New Integrable Canonical Structures in Two-dimensional Models},
 \href{http://www.sciencedirect.com/science/article/pii/0550321386903652}{Nucl. Phys. {\bf B269} (1986) 54.}

\bibitem{Maillet}
  J.~M.~Maillet,
  {\it Hamiltonian Structures for Integrable Classical Theories From Graded Kac-moody Algebras},
 \href{http://www.sciencedirect.com/science/article/pii/037026938691289X}{Phys. Lett. {\bf B167} (1986), 401.}


\bibitem{Ecker:1971xko}
G.~Ecker and J.~Honerkamp,
{\it Application of invariant renormalization to the nonlinear chiral invariant pion lagrangian in the one-loop approximation},
\href{https://www.sciencedirect.com/science/article/abs/pii/0550321371904688}{Nucl. Phys. \textbf{B35} (1971), 481}.


\bibitem{Friedan:1980jm}
D.~H.~Friedan,
{\it Nonlinear Models in Two + Epsilon Dimensions},
\href{https://www.sciencedirect.com/science/article/abs/pii/0003491685903847}{Annals Phys. \textbf{163} (1985), 318}.

\bibitem{Brezin:1990sk}
E.~Brezin and J.~Zinn-Justin,
{\it Fields, strings and critical phenomena. Proceedings}, Les Houches Lectures, 1989.

\bibitem{Tseytlin:1987bz}
A.~A.~Tseytlin,
{\it Conditions of Weyl Invariance of Two-dimensional $\sigma$ Model From Equations of Stationarity of 'Central Charge' Action},
\href{https://www.sciencedirect.com/science/article/pii/0370269387907702?via\%3Dihub}{Phys.\ Lett.\ {\bf B194} (1987), 63.}

\bibitem{Tseytlin:2006ak}
A.~A.~Tseytlin,
{\it On sigma model RG flow, 'central charge' action and Perelman's entropy},
Phys.\ Rev.\  {\bf D75} (2007), 064024,
\href{https://arxiv.org/abs/hep-th/0612296}{\tt hep-th/0612296.}

\bibitem{Sagkrioti:2018abh}
E.~Sagkrioti, K.~Sfetsos and K.~Siampos,
{\it Weyl anomaly and the $C$-function in $\lambda$-deformed CFTs},
\href{https://www.sciencedirect.com/science/article/pii/S0550321318303407?via\%3Dihub}{Nucl. Phys. \textbf{B938} (2019), 426},
\href{https://arxiv.org/abs/1810.04189}{\tt arXiv:1810.04189 [hep-th]}.



\bibitem{Petropoulos:2006py}
P.~M.~Petropoulos and K.~Sfetsos,
{\it Non-Abelian coset string backgrounds from asymptotic and initial data},
\href{https://iopscience.iop.org/article/10.1088/1126-6708/2007/04/033}{JHEP \textbf{04} (2007), 033},
\href{https://arxiv.org/abs/hep-th/0610055}{\tt hep-th/0610055}.

\bibitem{Itsios:2021wso}
G.~Itsios, K.~Sfetsos and K.~Siampos,
{\it Novel integrable interpolations},
\href{https://www.sciencedirect.com/science/article/pii/S0550321321002121?via\%3Dihub}{Nucl. Phys. \textbf{B971} (2021), 115515},
\href{https://arxiv.org/abs/2106.00032}{\tt arXiv:2106.00032 [hep-th]}.

\bibitem{Lozano:2011kb}
Y.~Lozano, E.~O Colgain, K.~Sfetsos and D.~C.~Thompson,
{\it Non-abelian T-duality, Ramond Fields and Coset Geometries},
\href{https://link.springer.com/article/10.1007/JHEP06(2011)106}{JHEP \textbf{06} (2011), 106},
\href{https://arxiv.org/abs/1104.5196}{{\tt arXiv:1104.5196 [hep-th]}}.

\bibitem{Delduc:2013fga}
F.~Delduc, M.~Magro and B.~Vicedo,
{\it On classical $q$-deformations of integrable sigma-models},
\href{https://link.springer.com/article/10.1007/JHEP11(2013)192}{JHEP \textbf{11} (2013), 192},
\href{https://arxiv.org/abs/1308.3581}{\tt arXiv:1308.3581 [hep-th]}.

\bibitem{Delduc:2013qra}
F.~Delduc, M.~Magro and B.~Vicedo,
{\it An integrable deformation of the $AdS_5 \times S^5$ superstring action},
\href{https://journals.aps.org/prl/abstract/10.1103/PhysRevLett.112.051601}{Phys. Rev. Lett. {\bf 112} (2014), 051601},
\href{https://arxiv.org/abs/1309.5850}{\tt arXiv:1309.5850 [hep-th]}.

 \bibitem{Vicedo:2015pna}
  B.~Vicedo,
  {\it Deformed integrable $\sigma$-models, classical $R$-matrices and classical exchange algebra on Drinfel'd doubles},
  \href{https://iopscience.iop.org/article/10.1088/1751-8113/48/35/355203}{J. Phys. A: Math. Theor. {\bf 48} (2015), 355203},
 \href{http://arxiv.org/abs/1504.06303}{\tt arXiv:1504.06303 [hep-th]}.

\bibitem{Hoare:2015gda}
  B.~Hoare and A.~A.~Tseytlin,
  {\it On integrable deformations of superstring sigma models related to $AdS_n \times S^n$ supercosets},
  \href{https://www.sciencedirect.com/science/article/pii/S0550321315002035?via\%3Dihub}{Nucl.\ Phys.\ {\bf B897} (2015), 448},
    \href{http://arxiv.org/abs/1504.07213}{\tt arXiv:1504.07213 [hep-th]}.
    

 \bibitem{Sfetsos:2015nya}
  K.~Sfetsos, K.~Siampos and D.~C.~Thompson,
 {\it Generalised integrable $\lambda$- and $\eta$-deformations and their relation},
  \href{https://linkinghub.elsevier.com/retrieve/pii/S0550321315003004}{Nucl. Phys. {\bf B899} (2015), 489},
  \href{http://arxiv.org/abs/1506.05784}{\tt arXiv:1506.05784 [hep-th]}.

\bibitem{Klimcik:2015gba}
C. Klim\v c\'\i k,
  {\it $\eta$ and $\lambda$ deformations as ${\cal E}$-models}, 
\href{https://www.sciencedirect.com/science/article/pii/S0550321315003302?via\%3Dihub}{Nucl.\ Phys.\ {\bf B900} (2015), 259},
\href{http://arxiv.org/abs/1508.05832}{\tt arXiv:1508.05832 [hep-th]}.


\bibitem{Klimcik:2016rov}
C.~Klim\v{c}\'\i{}k,
{\it Poisson\textendash{}Lie T-duals of the bi-Yang\textendash{}Baxter models},
\href{https://www.sciencedirect.com/science/article/pii/S0370269316303380?via\%3Dihub}{Phys. Lett. \textbf{B760} (2016), 345},
\href{https://arxiv.org/abs/1606.03016}{\tt arXiv:1606.03016 [hep-th]}.

\bibitem{Arutyunov:2015mqj}
G.~Arutyunov, S.~Frolov, B.~Hoare, R.~Roiban and A.~A.~Tseytlin,
{\it Scale invariance of the $\eta$-deformed $AdS_5\times S^5$ superstring, T-duality and modified type II equations},
\href{https://www.sciencedirect.com/science/article/pii/S0550321315004320?via\%3Dihub}{Nucl. Phys. \textbf{B903} (2016), 262},
\href{https://arxiv.org/abs/1511.05795}{\tt arXiv:1511.05795 [hep-th]}.

\bibitem{Itsios:2024gqr}
G.~Itsios,
{\it Type-II backgrounds from deformed coset CFTs},
\href{https://link.springer.com/article/10.1007/JHEP05(2025)095}{JHEP \textbf{05} (2025), 095},
\href{https://arxiv.org/abs/2411.11086}{\tt arXiv:2411.11086 [hep-th]}.

\end{thebibliography}

\end{document}